\documentclass[aps,prd,nofootinbib,floatfix,twocolumn]{revtex4-2}
\usepackage{amsmath,amsfonts,amssymb,color,graphicx,bm,hyperref,mathrsfs}
\usepackage{mathptmx}
\usepackage{physics}
\usepackage{framed}

\newcommand{\ie}{{\it i.e., }}

\begin{document}
\title{\textbf{
Can spacetime fluctuations generate entanglement between co-moving accelerated detectors?}}
\author{Dipankar Barman}
\email{dipankar1998@iitg.ac.in}
\author{Bibhas Ranjan Majhi}
\email{bibhas.majhi@iitg.ac.in}
\affiliation{Department of Physics, Indian Institute of Technology Guwahati, Guwahati 781039, Assam, India.}

\begin{abstract}

Recent studies [Class. Quant. Grav. 42, 03LT01 (2025); Phys. Rev. D 111, 045023 (2025)] indicate that in a nested sequence of Rindler wedges, vacuum of former Rindler frame appears to be thermally populated for an observer in shifted Rindler frame. Interestingly, this thermality is independent of shift parameter as long as it is non-zero and therefore arises even if the shift parameter is as small as Planck length. Building on this insight, we propose a set-up involving two atoms accelerating with identical acceleration. We find that if their Rindler frames (consequently their  trajectories) get infinitesimally separated, the atoms become entangled. Remarkably again, this entanglement, like the perceived thermality, is independent of the shift parameter, provided it is non-vanishing. Further we observe the vanishing of mutual information and discord. It implies the absence of both classical and non-classical correlations which are not related to entanglement. We investigate the dependence of entanglement on acceleration of the detectors. The present study indicates that the entanglement between two detectors, moving on the same Rindler wedge, is possible. Moreover, small spacetime fluctuations can lead to entanglement between detectors, moving along same classical trajectory. Hence we feel that such theoretical prediction has potential to probe the Planck length nature of spacetime.    

\end{abstract}
\maketitle	

\section{Introduction}

An intriguing prediction of quantum theory in the context of non-locality is quantum entanglement. Two quantum systems of interest can be correlated such that measurement over a subsystem's quantum state immediately influences state of other subsystem. Such quantum systems can be entangled even if they are spacelike separated. This phenomenon serves as a resource for various quantum information theoretic processes  {\it i.e.,} quantum teleportation \cite{Hotta:2008uk, Hotta:2009, Matson:2012aa, Frey:2014}, quantum cryptography \cite{PhysRevLett.67.661, Tittel:1998ja, Salart-2008, Yin:2020aa}, and computation \cite{Mooney:2019aa}, {\it etc}. This phenomenon remains very crucial even in the context of quantum field theory. In quantum field theory, vacuum fluctuations spontaneously create pairs of particle and antiparticle, which are entangled in general. Indeed, vacuum of any quantum field theory maximally violates of the Bell-CHSH inequality \cite{SUMMERS1985257, Summers:1987ze, 10.1063/1.527734, doi:10.1063/1.527733, RevModPhys.76.93}, affirming the entanglement content present in the vacuum. This entanglement structure of the vacuum also plays a key role behind the Hawking effect \cite{HAWKING:1974us, Hawking1975, Unruh:1976db, book:Birrell} and the Unruh effect \cite{Unruh:1976db, Unruh:1983ms, Crispino:2007eb, Takagi01031986, Higuchi:2017gcd}. 

In the seminal work \cite{VALENTINI1991321}, Valentini initially explored the feasibility of probing this vacuum entanglement employing two two-level atoms. These atoms, now widely referred to as Unruh-DeWitt (UDW) detectors \cite{Unruh:1976db, book:Birrell}, possess the capability to extract entanglement from the vacuum and subsequently become entangled. Consequently, the pioneering works by Reznik \cite{Reznik:2003uz, Reznik:2003mnx} established a systematic approach to investigate this phenomenon, which is further improved in
\cite{Koga:2018the, Ng:2018ilp}. The phenomenon of entanglement extraction from vacuum of field is now popularly known as {\it entanglement harvesting}. Entanglement harvesting is possible even if the detectors are causally disconnected and independent of the internal structure of the detectors. This process of swapping field entanglement to detectors turns out to be sensitive to the motion of the detectors \cite{Koga:2018the, Koga:2019fqh, Barman:2021bbw, Barman:2022xht, Wu:2024whx}, 
the nature of the background fields \cite{Pozas-Kerstjens:2016rsh, Perche:2021clp, Barman:2021bbw}, 
boundary conditions \cite{Liu:2020jaj, PhysRevD.108.085007}, 
the presence of black holes \cite{Menezes:2017oeb, Tjoa:2020eqh, PhysRevD.104.025001,  Barman:2021kwg, Barman:2023rhd}
 and other curved spacetimes \cite{ Kukita:2017etu,     Martin-Martinez:2015qwa, Barman:2021kwg,   K:2023oon, Ji:2024fcq, Lin:2024roh, Mayank:2025bkc}, 
 {\it etc}. Also if two detectors are initially entangled, they can lose entanglement due to vacuum fluctuations \cite{Barman:2022loh} and thermal bath \cite{Chowdhury:2021ieg}. In this regard, it may be mentioned that entanglement harvesting plays a major role to probe the quantum nature of gravity \cite{Nandi:2024zxp, Nandi:2024jyf, Dutta:2025bge}.
 
 Nature of entanglement harvesting also depends on the switching function of the interactions \cite{Pozas-Kerstjens:2017xjr, Koga:2018the, Koga:2019fqh, Henderson:2020ucx}. However, one should note that utilizing finite time interaction switching function introduces oscillatory transient effects, which can dominate over the original response of the UDW detectors \cite{Koga:2018the, Shallue:2025zto, Stargen:2025prb}. Thus, to extract the sole information of vacuum entanglements of quantum fields in the detectors one can consider eternal interaction between detectors and field, which allows one to avoid the spurious transient effects \cite{book:Birrell, Stargen:2025prb}. This provides a nice theoretical estimations about our main target of interest. However, in practical set-ups the \textit{infinte} time-interval is not needed. The environment of eternal interaction can be mimicked in laboratory by imposing the condition that the interaction time has to be much larger compared to the time scale associated with the detector's energy gap (e.g. see the discussion in \cite{Stargen:2025prb}). Therefore for the purpose of theoretical predictions it is legitimate to consider this simplified model which we adopt here. For such interaction time-periods in free Minkowski space, it is known that two accelerated detectors can get entangled if they are only in anti-parallel motion \cite{Koga:2019fqh, Barman:2021bbw, PhysRevD.108.085007}.  Conversely, when the detectors accelerate within the same Rindler wedge, they cannot become entangled.

 
In a recent study \cite{Lochan:2025mru}, the authors considered a nested sequence of Rindler frames with finite longitudinal shifts between them. They demonstrated that, for any two consecutive Rindler frames, the vacuum state of the preceding frame appears thermally populated to an observer in the next shifted frame. Remarkably, the expectation value of the number operator associated with the shifted Rindler observer, when evaluated in the vacuum of the earlier frame, is independent of the shift length $\ell$, as long as $\ell \neq 0$. However, this thermal behaviour vanishes when the shift is exactly zero. 
Building on this idea, a subsequent study \cite{PhysRevD.111.045023} utilized the Unruh-DeWitt (UDW) particle detector model to provide an operational framework for observing this effect. In their setup, two observers are considered accelerating in two subsequent Rindler frames, and the accelerated observer in the shifted Rindler frame probes the vacuum state of another observer in the preceding frame. When the detector is switched on for an infinite duration, its response is found to be thermal. Importantly, as in the earlier study \cite{Lochan:2025mru}, this thermal response is independent of the shift length, provided it is non-zero \footnote{Similar shift independent nature of expectation value of the number operator is also found in the following studies \cite{Gutti:2022xov, Nair:2024ryr}.}. 
As mentioned in \cite{Lochan:2025mru,PhysRevD.111.045023}, these insights have some intriguing implications. Since the Planck length ($\ell_p$) is widely regarded as the minimum physically meaningful length scale, one can take the shift length to be of the order of $\ell_p$ and still observe this emergent thermality between the frames. Thus, if a black hole's horizon shifts by the Planck scale length due to Hawking radiation, an exterior accelerating observer asymptoting to the former horizon perceives the new vacuum as thermally populated.  Similarly, suppose two accelerated observers initially follow the same trajectory but experience a relative shift of the order of the Planck length due to spacetime fluctuations, then also the preceding observer's vacuum appears to be thermal to the other observer. In that case, the shifted observer can perceive particles in the vacuum corresponding to the other observer. In this way, UDW detectors may capture the subtle imprints of Planck scale physics.

In this study, we consider a similar setup involving two Unruh-DeWitt (UDW) detectors, each undergoing uniform acceleration in successive Rindler wedges. It is well known that when two detectors follow such shifted accelerated trajectories while eternally interacting with the Minkowski vacuum; they cannot harvest entanglement (see Section III.B.3 of \cite{Koga:2019fqh}).  Motivated by this, we primarily aim to investigate whether these accelerated detectors can extract entanglement from the Rindler vacuum of the preceding Rindler wedge. Further, if the detectors become entangled, how does the amount of entanglement depend on the shift length? More importantly, whether this entanglement is independent of $\ell$ and exactly vanishes for $\ell = 0$? 
We find that entanglement harvesting is indeed possible in such a setup. Assuming identical parameter acceleration for both detectors, we observe that entanglement harvesting occurs for sufficiently large accelerations. Moreover, the amount of entanglement monotonically increases with acceleration. Interestingly, analogous to the detector response and the expectation value of the number operator, the harvested entanglement is independent of the shift $\ell$, as long as $\ell \neq 0$. When the shift is exactly zero, entanglement harvesting does not occur. 

We find present observations carry more robust indications towards the capturing the Planck length physics of spacetime. In the earlier studies \cite{Lochan:2025mru, PhysRevD.111.045023} it has been argued through the observation that the shifted detector will perceive the preceding Rindler vacuum to be thermally populated. However, apart from this the fluctuations of Minkowski vacuum  also provides excitation to this shifted detector (known as Unruh effect \cite{Unruh:1976db, Unruh:1983ms}). Therefore excitations can happen to the detector even if it is not being shifted from preceding accelerated observer. So both spacetime fluctuations as well as background Minkowski vacuum fluctuations promote excitation to the detector. Thus, from the observational point of view, it will become hard to figure out which vacuum is actually detected by the observer. Therefore, whether the observation is from the imprint of the Planck scale effect or from the Minkowski vacuum, can not be specified concretely. However, this ambiguity of observation does not arise through the study of entanglement harvesting. As shown in \cite{Koga:2019fqh}, if the observer detects the Minkowski vacuum fluctuations, there will be no entanglement irrespective of choice $\ell=0$ or $\ell\neq0$.
In contrast, our study demonstrates that entanglement harvesting is possible when the observers detect the Rindler vacuum only for $\ell\neq0$. Therefore, if any entanglement is observed, it will be precisely due to the fluctuations in the Rindler vacuum for non-zero $\ell$. Here again, we mention that $\ell$ can be taken in the order of $\ell_{p}$. Therefore,
 two accelerated detectors, moving along the same classical trajectory, if get entangled, then it must be due to a non-vanishing separation of their trajectories (i.e. the common Rindler horizon is get separated) at least of $\ell_{p}$ order due to small spacetime fluctuations. 
Thus, it seems that the entanglement harvesting can provide more robust and probably a unambiguous framework for exploring the signatures of the Planck length physics of spacetime.

The whole analysis is being done within the second order in perturbation of the system density matrix. We also study the mutual information \cite{book:nielsen, Simidzija:2018ddw, Barman:2021bbw} and quantum discord \cite{PhysRevLett.88.017901, Henderson_2001, PhysRevA.80.022108, PhysRevA.77.042303, PhysRevA.81.042105, PhysRevA.84.042313, Pal_2011, PhysRevA.88.014302, Yurischev_2015, YuGuo_2016, Lin:2024roh} for our system. It is observed that both the quantities vanish up to second order in perturbation for our specific set-up. It might be noted that these quantities are defined through classical concept where the classical inputs are being replaced by their quantum analogs. Mutual information carries information of both classical and quantum correlations, whereas quantum discord measures non-classical correlations, which are not necessarily related to quantum entanglement \cite{e23070797}. Hence vanishing of mutual information and discord while non-vanishing of entanglement imply absence of other quantum correlations unrelated to entanglement induced from the vacuum of one of the Rindler observers.



This paper is organized as follows. In Sec. \ref{Sec:2}, we discuss the framework for entanglement harvesting between two UDW detectors interacting with a minimally coupled, massless real scalar field through monopole coupling. This section discusses the mathematical description of the entanglement harvesting condition. In Sec. \ref{Sec3}, we discuss the trajectories and the Green’s functions of the two accelerated UDW detectors in nested Rindler frames. Subsequently, in Sec. \ref{Sec4}, we discuss the possibility of entanglement harvesting between the detectors in the nested Rindler frames. Also, the properties of the harvested entanglement are analyzed. Then we compare our results with the case where the shift is absent. 
In Sec. \ref{MI}, we calculate the mutual information and quantum discord for our system.
Finally, in Sec. \ref{Sec6}, we conclude with an overall discussion of the results. At the end we include three appendices to present details of the calculations.
 
\section{The model: framework for entanglement harvesting}\label{Sec:2}
Let us now briefly present our model of two UDW detectors interacting with background quantum field. Following the analysis presented in \cite{Koga:2018the,Ng:2018ilp}, the primary working formulas, valid up to the second-order perturbative expansion, will be summarised here. We consider two two-level Unruh-DeWitt detectors, denoted as $A$ and $B$, carried by the observers named Alice and Bob, respectively. For simplicity, we assume the detectors to be point-like and interacting with a massless real scalar field $\hat{\phi}(x)$ through monopole interaction. This interaction mimics the light-matter interactions like $\hat{\mathbf{p}}\cdot\hat{\mathbf{A}}$ (or  $\hat{\boldsymbol{\mu}}\cdot\hat{\mathbf{E}}$) when exchange of angular momentum is not concerned \cite{PhysRevD.97.105026}. The interaction action is given by
\begin{eqnarray}
\hat{S}_{I n t} &=&\sum_{j=A,B}\lambda_{j} \int_{-\infty}^{\infty}  d \tau_{j}\,\kappa_{j}\left(\tau_{j}\right) 
\hat{m}_{j}\left(\tau_{j}\right) \hat{\phi}\left(x_{j}\left(\tau_{j}\right)\right)\,,
\end{eqnarray}
where $\lambda_{j}$ is the coupling constant between the $j^{th}$ detector ($j=A,B$) and the scalar field. $\kappa_{j}\left(\tau_{j}\right)$ and $\tau_{j}$ are the interaction switching function and proper time for the $j^{th}$ detector, respectively. $\hat{\phi}$ is the field operator corresponding to background real scalar field. For simplicity of the model and analytically handling the computations, we will use $\kappa(\tau)=1$ for eternal interaction between the detectors and field as used in \cite{Koga:2018the, Koga:2019fqh, Barman:2021bbw, Barman:2021kwg, Barman:2022xht, Barman:2022loh, PhysRevD.108.085007}. This helps to suppress spurious transient effects \cite{book:Birrell, Stargen:2025prb}. However, in practical set-ups this \textit{infinte} time-interval is not needed, it has to be much larger compared to the time scale associated with the detector's energy gap. The monopole operator of the detector's are given as 
\begin{equation}\label{M}
\hat{m}_{j}(\tau_{j})=e^{i\hat{H}_{j}\tau_{j}}(|e_{j}\rangle\langle g_{j}|+|g_{j}\rangle\langle e_{j}|)e^{-i\hat{H}_{j}\tau_{j}}\,.
\end{equation}
 Here $|g_{j}\rangle$ and $|e_{j}\rangle$ are the ground and exited states of the $j^{th}$ detector, respectively.

The initial state of the composite system is taken to be 
$|in\rangle=|\Psi\rangle|g_{A}\rangle|g_{B}\rangle$, where $|\Psi\rangle$ is the field vacuum. The final state of the system in the asymptotic future can be obtained as $|out\rangle=T\{e^{iS_{int}}\}|in\rangle$, where $T$ denotes the time order product. One can get the reduced density matrix for the detectors $\rho_{AB}$ by tracing out the field degrees of freedom from the final total density matrix, which in the basis of 
$\{|e_{A}\rangle|e_{B} \rangle,|e_{A} \rangle|g_{B}\rangle, 
|g_{A}\rangle|e_{B} \rangle,|g_{A}\rangle|g_{B}\rangle\}$ is expressed as \cite{Koga:2018the}
%
\begin{equation}\label{eq:detector-density-matrix}
\rho_{AB}=\begin{pmatrix}
	0			&0					&0				&\lambda^{2}\mathcal{E}\\
	0			&\lambda^{2}\mathcal{P}_{A}		&\lambda{2}\mathcal{P}_{AB}	&0\\
	0			&\lambda^{2}\mathcal{P}^{\star}_{AB}	&\lambda^{2}\mathcal{P}_{B}	&0\\	
\lambda^{2}\mathcal{E}^{\star}	&0					&0				&1-\lambda^{2}\mathcal{P}_{A}-\lambda^{2}\mathcal{P}_{B}
\end{pmatrix}+O(\lambda^{4}).
\end{equation}
Here for identical detector atoms one can take $\lambda_{A}=\lambda_{B}=\lambda$. It may be pointed that $\mathcal{P}_j$ can be identified as the individual detector's transition probability, whereas in literature $\mathcal{E}$ is usually called as entangling term. The structure of the density matrix depends on the choice of the initial detectors' state and the monopole operator considered. For our particular monopole operator given in Eq. (\ref{M}), expressions for the detectors' density matrix elements are
\begin{eqnarray}\label{IExpressions}
\mathcal{P}_{j}&=&\int_{-\infty}^{\infty}\int_{-\infty}^{\infty}{d\tau_{j}d\tau'_{j}}\,e^{
-i\Omega(\tau_{j}-\tau'_{j})}\,G_{W}(x_{j},x'_{j})\,,
\nonumber\\ 
\mathcal{E}&=&-\int_{-\infty}^{\infty}\int_{-\infty}^{\infty}{d\tau_{B}d\tau'_{A}}\,
e^{i\Omega(\tau'_{A}+\tau_{B})}\,iG_{F}(x_{B},x'_{A})\,,
\nonumber\\
\mathcal{P}_{AB}&=&\int_{-\infty}^{\infty}\int_{-\infty}^{\infty}{d\tau_{B}d\tau'_{A}}\,e^{i\Omega(\tau'_{A
}-\tau_{B})}\,G_{W}(x_{B},x'_{A})\,.
 \end{eqnarray}
%
Here $\Omega$ is the energy gap of the detectors. The quantities $G_{W}(x_{i},x'_{j}),\,G_{F}(x_{i},x'_{j})$ are respectively the positive frequency Wightman function and the Feynman propagator, defined as
\begin{eqnarray}
G_{W}(x_{i},x'_{j})&=&\langle\Psi|\hat{\phi}(x_{i})\hat{\phi}(x_{j}')|\Psi\rangle\,,\nonumber\\
iG_{F}(x_{i},x'_{j})&=&\langle\Psi|T\{\hat{\phi}(x_{i})\hat{\phi}(x_{j}')\}|\Psi\rangle
\,~.
\end{eqnarray}

Since our system is a bipartite system, any negative  eigenvalue of the partial transposition of the reduced density matrix (see Eq.(\ref{eq:detector-density-matrix})) confirms entanglement between the detectors \cite{Peres:1996dw, Horodecki:1996nc}. The absolute value of sum of all negative eigenvalues is known as the {\it negativity}, a measure of entanglement. For our density matrix, there will be a negative eigenvalue if the following condition is satisfied \cite{Koga:2018the, Koga:2019fqh}
\begin{equation}\label{eq:cond-entanglement}
 \mathcal{P}_{A}\,\mathcal{P}_{B}<|\mathcal{E}|^2~.
\end{equation}%
Once the above condition is satisfied, one may 
study various measures to quantify the harvested entanglement. In this regard, a convenient entanglement measures is the {\it concurrence} \cite{Bennett:1996gf, Hill:1997pfa, Wootters:1997id}, which is very useful for estimating the entanglement of formation $E_{F}(\rho_{AB})$ (see \cite{Bennett:1996gf, Hill:1997pfa, Wootters:1997id, Koga:2018the}). For our two qubits system this quantity is obtained as  \cite{Koga:2018the} 
\begin{eqnarray}\label{eq:concurrence-gen-exp}
&&\mathnormal{C}(\rho_{AB})= {\it max}\{0,\,2\lambda^2 \left(|\mathcal{E}|-\sqrt{\mathcal{P}_{A}\mathcal{P}_{B}}\right)\}\,.
\end{eqnarray}
Studying this quantity one can understand the influence of the background spacetime and motions of detectors on entanglement between two detectors.

The entirety of classical and quantum correlations, between the detectors $A$ and $B$ is quantified by mutual information $\mathcal{M}$, defined as \cite{book:nielsen}
\begin{equation}\label{MI}\begin{aligned}
\mathcal{M}(\rho_{AB})=S(\rho_{A})+S(\rho_{B})-S(\rho_{AB})\,,
\end{aligned}\end{equation}
where $S(\rho)=-\text{Tr}[\rho\log\rho]$ is the von Neumann entropy. This definition is an extension of the mutual information from classical information theory, where the Shannon entropy is replaced by the von Neumann entropy. For our present setup we find \cite{Simidzija:2018ddw, Barman:2021bbw}
\begin{equation}\label{eq:MI-explicit}\begin{aligned}
 \mathcal{M}(\rho_{AB}) &= \lambda^2\big[\mathcal{P}_{+}\ln{\mathcal{P}_{+}} + \mathcal{P}_{-}\ln{\mathcal{P}_{-}} - 
\mathcal{P}_{A}\ln{\mathcal{P}_{A}}\\&~~~~~~~~~~~ - \mathcal{P}_{B}\ln{\mathcal{P}_{B}}\big] + \mathcal{O}(\lambda^4)~,
\end{aligned}\end{equation}
where, the quantities $\mathcal{P}_{\pm}$ are given by
\begin{equation}\label{eq:P-pm}
 \mathcal{P}_{\pm} = \frac{1}{2} \Big[ \mathcal{P}_{A}+\mathcal{P}_{B}\pm \sqrt{(\mathcal{P}_{A}-\mathcal{P}_{B})^2+4|\mathcal{P}_{AB}|^2} 
\Big]~.
\end{equation}
 On the other hand, an alternative expression of quantum mutual information can be defined by introducing a projective measurement ($B_{k}$) over subsystem $B$ as \cite{PhysRevLett.88.017901, Henderson_2001, PhysRevA.80.022108, PhysRevA.77.042303, PhysRevA.81.042105, PhysRevA.84.042313, Pal_2011, PhysRevA.88.014302, Yurischev_2015, YuGuo_2016, Lin:2024roh}
\begin{equation}\label{DefJ}\begin{aligned}
\mathcal{J}(\rho_{AB}) =S(\rho_{A})-S(\rho_{AB}|\{B_{k}\})\,,
\end{aligned}\end{equation}
where $S(\rho_{AB}|\{B_{k}\})$ is quantum analog conditional entropy. Maximum of $\mathcal{J}(\rho_{AB})$ provides measure of the classical correlations \cite{Henderson_2001}. The variation between these two definitions of mutual informations is known as quantum discord \cite{PhysRevLett.88.017901, Henderson_2001, PhysRevA.80.022108, PhysRevA.77.042303, PhysRevA.81.042105, PhysRevA.84.042313, Pal_2011, PhysRevA.88.014302, Yurischev_2015, YuGuo_2016, Lin:2024roh}
\begin{equation}\label{Disc}\begin{aligned}
\mathcal{D}(\rho_{AB})=\mathcal{M}(\rho_{AB})-\text{Max}_{\{B_{k}\}}\mathcal{J}(\rho_{AB})\,.
\end{aligned}\end{equation}
In general, a quantum measurement change the original state, leading to a variation between two different quantum analogs of the identical classical mutual information. 
If both $\mathcal{M}$ and $\mathcal{J}$ vanishes (so $\mathcal{D}$)  in a system, but concurrence is non-vanishing, that would imply absence of quantum correlations unrelated to quantum entanglement \cite{e23070797}.

To obtain concurrence, mutual information and quantum discord, we need to evaluate the quantities $\mathcal{P}_{j}$, $\mathcal{P}_{AB}$ and $\mathcal{E}$. For that we will require the relevant green's functions. Let us calculate these greens functions in the next section.

\section{Accelerated observers in Nested Rindler frames}\label{Sec3}
We confine the present analysis in $(1+1)$ spacetime dimensions (like as adopted in \cite{Lochan:2025mru,PhysRevD.111.045023}).
Let us consider Alice is moving in the $R_{1}$ Rindler frame $(\eta_{1},\xi_{1})$ and Bob in the $R_{0}$ Rindler frame $(\eta_{0},\xi_{0})$. These Rindler frames are associated with two shifted Minkowski frames $M_{1}~(t',x')$ and $M_{0}~(t,x)$, respectively.  The relation between the Rindler and Minkowski co-ordinates are given as follows:
\begin{equation}\label{Relnxp}\begin{aligned}
\text { Between } R_0 \text { and } M_0: t & =\frac{1}{a_0} e^{a_0 \xi_0} \sinh \left(a_0 \eta_0\right)\,, \\
x & =\frac{1}{a_0} e^{a_0 \xi_0} \cosh \left(a_0 \eta_0\right)\,; \\
\text { Between } R_1 \text { and } M_1: t^{\prime} & =\frac{1}{a_1} e^{a_1 \xi_1} \sinh \left(a_1 \eta_1\right)\,, \\
x^{\prime} & =\frac{1}{a_1} e^{a_1 \xi_1} \cosh \left(a_1 \eta_1\right)\,~.
\end{aligned}
\end{equation}
Here the Minkowski frames are related as 
\begin{equation}\label{Relnxpx}\begin{aligned}
x^{\prime}&=x-\ell\,,\\
t'&=t\,.
\end{aligned}\end{equation}
\begin{figure}[h]
	\centering
	\footnotesize
\includegraphics[width=0.43\textwidth]{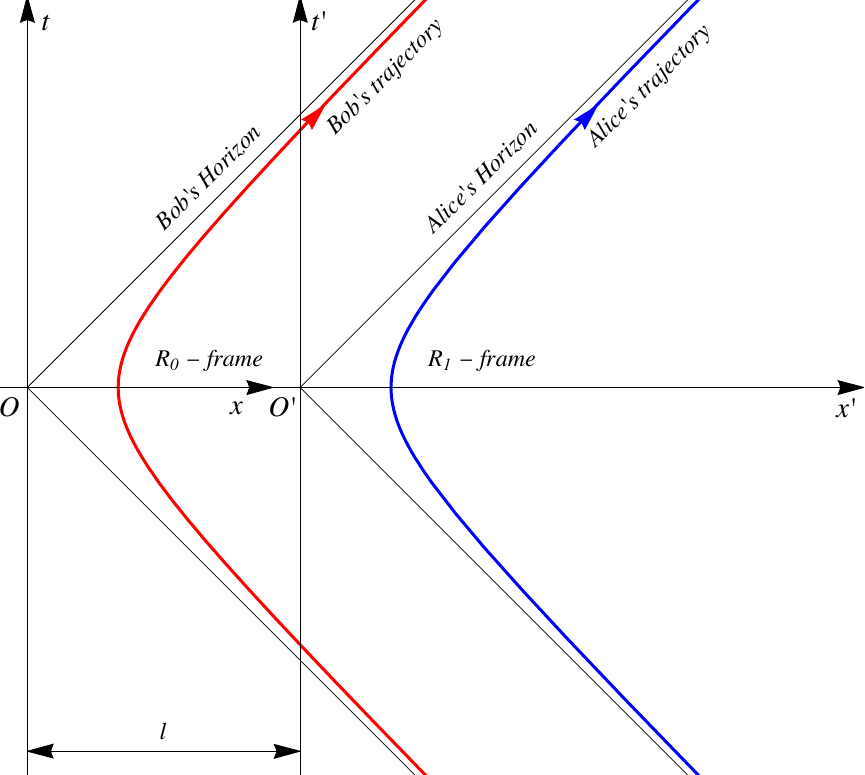}\\
\caption{Nested Rindler frames with observers Alice and Bob.
}
	\label{fig:figures0yx}
\end{figure}

For our study, we consider massless real scaler fields in $R_{0}$-frame. In $(1+1)$-dimension, the solutions of the Klein–Gordon equation (ingoing and outgoing) are given as \cite{book:Birrell}
\begin{equation}\label{UmodeR0x}
\begin{aligned}
u_{\omega_{0}}=\frac{e^{-i\omega_{0}{u}_{0}}}{\sqrt{4\pi\omega_{0}}}\,,\\
u_{-\omega_{0}}=\frac{e^{-i\omega_{0}{v}_{0}}}{\sqrt{4\pi\omega_{0}}}\,.
\end{aligned}
\end{equation}
Since we consider Alice and Bob present in the $R_{1}$ and $R_{0}$ frame, respectively. To calculate $\mathcal{P}_{B}$, we can directly use these field modes to evaluate the Wightman propagator. However, for calculating $\mathcal{P}_{A}$, we need to express these field modes in terms of $R_{1}$ coordinates. For evaluation of the entangling term $\mathcal{E}$, we require the Feynman propagator which involves the coordinates of both Alice and Bob. For that one need to express Alice's contribution in the Feynman propagator in $R_{1}$ coordinates. For these reasons, we need to express these field modes in terms of $R_{1}$ coordinates as well. Using the relations in Eq. (\ref{Relnxp}) and (\ref{Relnxpx}), one can express mode functions of $R_{0}$ frame in terms of $R_{1}$ coordinates as 
\begin{equation}\label{UmodeR0}
\begin{aligned}
u_{\omega_{0}}=&
\left( \frac{a_{0}}{a_{1}}\right)^{\frac{i\omega_{0}}{a_{0}}} \frac{e^{-\frac{i\omega_{0}}{a_{0}}a_{1}{u}_{1}}}{\sqrt{4\pi\omega_{0}}}(1+a_{1}\ell e^{a_{1}{u}_{1}})^{\frac{i\omega_{0}}{a_{0}}}\,,\\
u_{-\omega_{0}}=&
\left( \frac{a_{0}}{a_{1}}\right)^{-\frac{i\omega_{0}}{a_{0}}} \frac{e^{-\frac{i\omega_{0}}{a_{0}}a_{1}{v}_{1}}}{\sqrt{4\pi\omega_{0}}}(1+a_{1}\ell e^{-a_{1}{v}_{1}})^{-\frac{i\omega_{0}}{a_{0}}}\,;
\end{aligned}
\end{equation}
where, $({u}_0,{v}_0)=(\eta_{0}-\xi_{0},\eta_{0}+\xi_{0})$ and $({u}_1,{v}_1)=(\eta_{1}-\xi_{1},\eta_{1}+\xi_{1})$. The detailed calculation is given in Appendix \ref{ModesC}.

In $R_{0}$ frame, one can write the massless real scaler field as
\begin{equation}\label{}\begin{aligned}
\hat{\phi}({u}_{0},{v}_{0})&=\int{d\omega_{0}}[\hat{a}_{\omega_{0}}u_{\omega_{0}}+\hat{a}_{-\omega_{0}}u_{-\omega_{0}}\\&~~~~~~~~+
	\hat{a}^{\dagger}_{\omega_{0}}u^{\star}_{\omega_{0}}+
	\hat{a}^{\dagger}_{-\omega_{0}}u^{\star}_{-\omega_{0}}]\,,
\end{aligned}\end{equation}
with $\hat{a}_{\pm\omega_{0}}$ are annihilation operators. The field vacuum ($|\Psi\rangle
\,=\ket{0_{R_{0}}}$) is defined as $\hat{a}_{\pm\omega_{0}}\ket{0_{R_{0}}}=0$.  
Then the positive frequency Wightman function can be written as
\begin{equation}\label{ }\begin{aligned}
G_{W}(\eta_{0},\eta_{0}')&=  \bra{0_{R_{0}}}\hat{\phi}(\eta_{0})\hat{\phi}(\eta_{0}') \ket{0_{R_{0}}}
\\&=
\int{d\omega_{0}}[u_{\omega_{0}}u^{\prime\star}_{\omega_{0}}+u_{-\omega_{0}}u^{\prime\star}_{-\omega_{0}}]\,.
\end{aligned}
\end{equation}
To calculate the Wightman propagators consisting coordinates of both observers (particularly when the attention is on the cross correlator), we assign the 
primed modes to Alice and therefore express this primed modes in terms of $R_{1}$ coordinates. We assign the non-primed modes to Bob and keep this modes in terms of $R_{0}$ coordinates. Then using Eq.  (\ref{UmodeR0x}) and (\ref{UmodeR0}), we can rewrite this Green's function as
\begin{equation}\label{GR0}\begin{aligned}
	G_{W}(\eta_{0},\eta_{1})&\\	=\int\frac{d\omega_{0}}{4\pi\omega_{0}}&\left(e^{-i\omega_{0}{u}_{0}}
\left( \frac{a_{0}}{a_{1}}\right)^{-\frac{i\omega_{0}}{a_{0}}}e^{\frac{i\omega_{0}}{a_{0}}a_{1}{u}_{1}}(1+a_{1}\ell e^{a_{1}{u}_{1}})^{-\frac{i\omega_{0}}{a_{0}}}\right.
\\+&
\left.	e^{-i\omega_{0}{v}_{0}}
	\left( \frac{a_{0}}{a_{1}}\right)^{\frac{i\omega_{0}}{a_{0}}} e^{\frac{i\omega_{0}}{a_{0}}a_{1}{v}_{1}}(1+a_{1}\ell e^{-a_{1}{v}_{1}})^{\frac{i\omega_{0}}{a_{0}}}\right)\,.
	\end{aligned}\end{equation}
	Here we can set $\xi_{j}=0$. Then we have $\eta_{j}=\tau_{j}$, where $\tau_{j}$ is proper time of $j^{th}$ detector ($j=0,1$). Hence $G_{W}(\eta_{0},\eta_{1})$ can be written as $G_{W}(\tau_{0},\tau_{1})$.

\section{Entanglement Harvesting}\label{Sec4}
In this case, we consider both detectors are observing $R_{0}$-vacuum. Here Bob's detector (in $R_{0}$ frame) sees its in own vacuum. Thus the detector $B$'s transition probability is zero; i.e. $\mathcal{P}_{B}=0$. On the other hand, detector $A$ perceives $R_{0}$-vacuum as a thermal bath \cite{PhysRevD.111.045023} as
\begin{equation}\label{PAl}
\mathcal{P}_{A}=\frac{1}{a_{1}\Omega}\left[\frac{\pi\delta(0)-\log\frac{\Omega}{a_{1}}}{e^{\frac{2\pi\Omega}{a_{1}}}-1}-
\sum_{n=1}^{\infty}\Gamma\left[0,\frac{2\pi n\Omega}{a_{1}}\right]
\right]\,.
\end{equation}
However, to calculate the concurrence, we need to evaluate one of the quantities, like $\sqrt{\mathcal{P}_{A}\mathcal{P}_{B}}$ (see Eq. (\ref{eq:concurrence-gen-exp})), which turns out to be vanishing. Hence any non-zero value of $\mathcal{E}$ implies entanglement between the detectors.

Let us check the quantity $|\mathcal{E}|$. For our specified study we need to consider $a_{0}=a_{1}$, {\it i.e.,} equal parameter acceleration for both detectors. Thus we will require the Feynman propagator. The Feynman propagator can be expressed as
\begin{equation}\label{GF0}\begin{aligned}
iG_{F}(\tau_{0},\tau_{1})&=\theta(\tau_{0}-\tau_{1})G_{W}(\tau_{0},\tau_{1})+\theta(\tau_{1}-\tau_{0})G_{W}(\tau_{1},\tau_{0})\,,
\end{aligned}\end{equation}
where, $G_{W}(\tau_{0},\tau_{1})$ is given in Eq. (\ref{GR0}). Using this propagator we can write $\mathcal{E}$ in Eq. (\ref{IExpressions}) as

 \begin{equation}\label{EE0}\begin{aligned}
\mathcal{E}&=-
\int_{-\infty}^{\infty}d\tau_{1}\int_{\tau_{1}}^{\infty}d\tau_{0}\,e^{i\Omega(\tau_{0}+\tau_{1})}G_{W}(\tau_{0},\tau_{1})\\
-&\int_{-\infty}^{\infty}d\tau_{1}\int_{-\infty}^{\tau_{1}}d\tau_{0}\,e^{i\Omega(\tau_{0}+\tau_{1})}G_{W}(\tau_{1},\tau_{0})\,.
\end{aligned}\end{equation}
After evaluating $\tau_{0}$ and then $\tau_{1}$-integrals (see Appendix \ref{CalcR0}), we obtain
\begin{equation}\label{EE1x}\begin{aligned}
\mathcal{E}&=-\int_{-\infty}^{\infty}\frac{d\omega_{0}}{4\pi\omega_{0}}
\frac{i(a_{1}\ell)^{-i(\frac{2\Omega}{a_{1}})}B[i(\frac{2\Omega}{a_{1}}),\nu
-i(\frac{2\Omega-\omega_{0}}{a_{1}})]}{a_{1}(\Omega-\omega_{0}+i\epsilon)}
\\-&
\int_{-\infty}^{\infty}\frac{d\omega_{0}}{4\pi\omega_{0}}\frac{i(a_{1}\ell)^{+i(\frac{2\Omega}{a_{1}})}
B[-i(\frac{2\Omega}{a_{1}}),\nu+i(\frac{2\Omega-\omega_{0}}{a_{1}})]}{a_{1}(\Omega-\omega_{0}+i\epsilon)}\,.
\end{aligned}\end{equation}
Then performing these integrations in complex $\omega_{0}$-plane (see Appendix \ref{CalcR0}), we obtain the concurrence for this system as
\begin{equation}\label{EEF}
\mathnormal{C}=2\lambda^{2}|\mathcal{E}|=\frac{\lambda^{2}}{2\Omega^{2}}\sqrt{\frac{\frac{2\pi\Omega}{a_{1}}}{\sinh\frac{2\pi\Omega}{a_{1}}}}\,.
\end{equation}

 There are few interesting features of the above measured entanglement. First of all we find that two Rindler observers harvest entanglement when they are separated. However the entanglement is due to fluctuation of preceding Rindler observer's vacuum.
 Moreover, the concurrence turns out to be independent of $\ell$ with $\ell\neq 0$.   
 The $\ell = 0$ case needs to be diagnosed separately.
 For $\ell=0$ (with $a_{0}=a_{1}$), the Green's function in Eq. (\ref{GR0}) becomes
 \begin{equation}\label{GR00}\begin{aligned}
	G_{W}(\eta_{0},\eta_{1})&=\int\frac{d\omega_{0}}{4\pi\omega_{0}}[e^{-i\omega_{0}\tilde{u}_{0}}
e^{i\omega_{0}\tilde{u}_{1}}
+
	e^{-i\omega_{0}\tilde{v}_{0}}
e^{i\omega_{0}\tilde{v}_{1}}]
\\&=
\int\frac{d\omega_{0}}{2\pi\omega_{0}} e^{-i\omega_{0}(\tau_{0}-\tau_{1})}
\,.
	\end{aligned}\end{equation}
	Here $\tau_{j}=\eta_{j}$ as we took $\xi_{j}=0~(j=A,B)$. Using this Green's function in Eq. (\ref{EE0}) and performing $\tau_{0}$ integral, we obtain 
\begin{equation}\label{EE00}\begin{aligned}
\mathcal{E}&=
\int\frac{d\omega_{0}}{2\pi\omega_{0}}\int_{-\infty}^{\infty}d\tau_{1}e^{i(\Omega+\omega_{0})\tau_{1}}\frac{e^{i(\Omega-\omega_{0})\tau_{1}}}{i(\Omega-\omega_{0})}\\&
-\int\frac{d\omega_{0}}{2\pi\omega_{0}}\int_{-\infty}^{\infty}d\tau_{1}e^{i(\Omega-\omega_{0})\tau_{1}}\frac{e^{i(\Omega+\omega_{0})\tau_{1}}}{i(\Omega+\omega_{0})}\,.
\end{aligned}\end{equation}
After performing the integration on $\tau_{1}$, each term yields a Dirac delta function $\delta(2\Omega)$. Since we have $\Omega>0$ the entangling term becomes {\it zero}.
Interestingly, the same features were observed in the transition probability of $A$-detector (see Eq. (\ref{PAl})) and also in the expectation value of the number operator of $A$ with respect to vacuum of $B$ (see \cite{Lochan:2025mru, PhysRevD.111.045023} for details).

\begin{figure}[h]
	\centering
	\footnotesize
\includegraphics[width=0.43\textwidth]{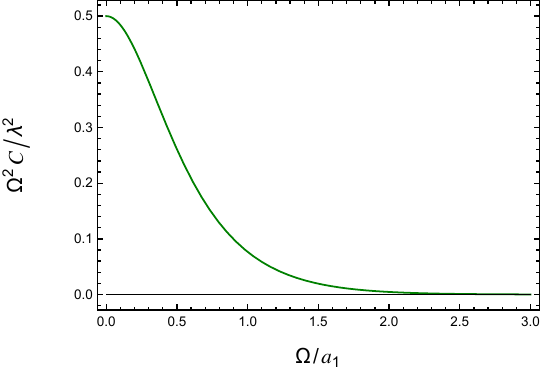}\\
\caption{We plotted dimensionless concurrence $\Omega^{2}\mathnormal{C}/\lambda^{2}$ with respect to $\Omega/a$.
}
	\label{fig:figures1yx}
\end{figure}

In the Fig. \ref{fig:figures1yx} we plotted the dimensionless concurrence $\Omega^{2}\mathnormal{C}/\lambda^{2}$ with respect to the dimensionless inverse acceleration $\Omega/a_{1}$. Here we observe that for smaller acceleration values, the quantity $\Omega^{2}\mathnormal{C}/\lambda^{2}$ is almost negligible. For sufficiently large accelerations, dimensionless concurrence monotonically increases with the accelerations. 
Also, towards the large accelerations, the increase of concurrence is very high compared to the moderate acceleration values.
Therefore such entanglement harvesting is feasible at large acceleration range. This analysis predicts a very interesting phenomenon. Suppose we start with two accelerated detectors, moving along same trajectory. If they maintain the same trajectory, then they will not be entangled. However a small fluctuation in the spacetime may kick them apart from each other by a small amount. Then they will face their individual Rindler horizons which are at a very small distance apart. In that case, our analysis predicts that the detectors must be entangled. However, contrary to transition to higher state in $A$ from both Minkowski vacuum as well as that of $B$, the entanglement is not possible from their common Minkowski vacuum. The latter one is possible only from the vacuum of $B$ and also when they are separated at least by any non-vanishing shift. Since the minimum length scale can be Planck length, the observation of entanglement between the atoms will carry the features of Planck scale physics of spacetime. Hence the entanglement harvesting phenomenon can be more reliable event than the transition of the detector to illuminate the physics at the Planck scale.

\section{Mutual information and quantum discord}\label{MI}
From Eq. (\ref{eq:MI-explicit}) and (\ref{eq:P-pm}) it is observed that mutual information for our system can be estimated by evaluating the quantities $\mathcal{P}_{j}~(j=A,\,B)$ and $\mathcal{P}_{AB}$. Since $\mathcal{P}_{j}$ is already known to us (note that $\mathcal{P}_{A}$ is given by (\ref{PAl}) and $\mathcal{P}_{B}=0$), we only need to evaluate $\mathcal{P}_{AB}$. Substituting Eq. (\ref{GR0}) in Eq. (\ref{IExpressions}) we find
\begin{equation}\label{PAB}\begin{aligned}
\mathcal{P}_{AB}&=\int_{0}^{\infty}\frac{d\omega_{0}}{4\pi\omega_{0}}\left\{\int_{-\infty}^{\infty}{d\tau_{0}}e^{-i\Omega\tau_{0}}e^{-i\omega_{0}\tau_{0}}\right\}\Bigg\{
\int_{-\infty}^{\infty}{d\tau_{1}}\,e^{i\Omega\tau_{1}}
\\&
\left[\left( \frac{a_{0}}{a_{1}}\right)^{-\frac{i\omega_{0}}{a_{0}}}e^{\frac{i\omega_{0}}{a_{0}}a_{1}{\tau}_{1}}(1+a_{1}\ell e^{a_{1}{\tau}_{1}})^{-\frac{i\omega_{0}}{a_{0}}}\right.
\\+&
\left.	\left( \frac{a_{0}}{a_{1}}\right)^{\frac{i\omega_{0}}{a_{0}}} e^{\frac{i\omega_{0}}{a_{0}}a_{1}{\tau}_{1}}(1+a_{1}\ell e^{-a_{1}{\tau}_{1}})^{\frac{i\omega_{0}}{a_{0}}}\right]\Bigg\}\,.
	\end{aligned}\end{equation}
Here evaluating the $\tau_{0}$-integral, one obtain $2\pi\delta(\Omega+\omega_{0})$, which vanishes as both $\Omega$ and $\omega_{0}$ are greater than zero. This is expected as the satisfaction of the positivity of the density matrix requires $\mathcal{P}_{A}\mathcal{P}_{B}\geq|\mathcal{P}_{AB}|^{2}$  \cite{Koga:2018the} where we have $\mathcal{P}_{B}=0$ for our system. Hence we can conclude that mutual information for our system is vanishing up to $\mathcal{O}(\lambda^{2})$.

On the other hand, the conditional entropy $S(\rho_{AB}|\{B_{k}\})$ (evaluated in Appendix \ref{MIQD}) is independent of minimization over $\{B_{k}\}$ up to $\mathcal{O}(\lambda^{2})$. In fact it is equal to $S(\rho_{A})$. Hence quantum discord also vanishes for our system (see Eq. (\ref{Disc})).

\section{Conclusion}\label{Sec6}
 In this study, following the setup proposed in \cite{Lochan:2025mru, PhysRevD.111.045023}, we investigate the possibility of entanglement harvesting between two uniformly accelerated Unruh-DeWitt (UDW) detectors. These earlier works demonstrated that a detector in a shifted Rindler frame perceives the Rindler vacuum $\ket{0_{R_{0}}}$ as thermally populated. Since, this thermality is independent of the shift length $\ell$, as long as $\ell \neq 0$, and vanishes when the shift is exactly zero. This allows one to consider $\ell$ as small as the Planck length $\ell_p$ and still observe the same thermal behavior. This observation suggest that two accelerated observers who initially follow the same trajectory can experience a $\ell_{p}$-order relative shift in the trajectories due to spacetime fluctuations. In that case, the shifted observer will perceive particles in the former vacuum $\ket{0_{R_{0}}}$. In this way, UDW detectors can capture the subtle imprints of Planck scale physics. However, instead of detecting the Rindler vacuum if the observer detects the global Minkowski vacuum, identical thermality will be detected. Also it is very hard to isolate our detectors from the influence of Minkowski vacuum. Consequently, the observation can be ambiguous as it remains unclear whether the thermality is a genuine signature of Planck-scale effects or simply a manifestation of the Unruh effect in Minkowski space.

Motivated by these insights, we investigated the possibility of entanglement harvesting when the detectors are coupled to the Rindler vacuum of $B$. Our goal was to determine whether the detectors can be entangled in such a setup and, if so, whether the harvested entanglement depend on the shift length $\ell$, analogous to the detector response. In particular, we focused on whether this framework could serve as a probe of Planck-scale physics. Our findings suggest that two accelerated observers, initially on identical trajectory but got separated by a Planck-scale shift due to spacetime fluctuations, can become entangled through the Rindler vacuum. The entanglement, quantified by concurrence, increases monotonically with the acceleration. Remarkably, similar to the thermal response observed in earlier studies, the concurrence is independent of the shift $\ell$, as long as $\ell \neq 0$. This implies that even when $\ell$ is taken to be of the order of the Planck length $\ell_p$, the same amount of entanglement is harvested. Importantly, unlike in the case of detector response -- where thermality could arise from either the Rindler or Minkowski vacuum, the entanglement harvesting scenario offers a clearer interpretation. It is well established that when detectors are eternally coupled to the Minkowski vacuum, they cannot become entangled \cite{Koga:2019fqh}. Therefore, any observed entanglement must be originated from fluctuations in the Rindler vacuum $\ket{0_{R_0}}$ alone along with a minimum value of $\ell \sim \ell_p$. This makes the phenomenon a more robust probe of Planck-scale effects. 

We have also investigated the quantities: mutual information and quantum discord. Our perturbative approach reveals that both mutual information and quantum discord vanishes for our system up to the $\lambda^{2}$-order. 
That implies absence of any classical and quantum  correlations in the system rather than quantum entanglement from field vacuum \cite{e23070797}.

As a final comment we must mention that the present implications are completely based on theoretical ground. Since we need very high acceleration to probe these physics, the practical feasibility within the current technology is out of reach. Nonetheless this theoretical study carries importance in predicting small scale properties of spacetime. Therefore it builds an important step to illuminate such physics. We feel that based on this study people will be able find meaningful idea to know the quantum physics of spacetime in future. Also the present analysis has been done in $(1+1)$ spaetime dimensions to avoid mathematical complexities. However $(3+1)$ dimensional analysis will be more interesting. We leave this for our future study.

\vskip 3mm
\begin{acknowledgments}
{\it Acknowledgments.}-- DB would like to acknowledge Ministry of Education, Government of India for
providing financial support for his research via the PMRF May 2021 scheme. The research of BRM is partially supported by a START-UP RESEARCH GRANT (No. SG/PHY/P/BRM/01) from the Indian Institute of Technology Guwahati, India.

\end{acknowledgments}


\begin{widetext}
\section*{Appendices}
\begin{appendix}
\numberwithin{equation}{section}
\section{Mode functions in Eq. (\ref{UmodeR0x}) in terms of $R_{1}$ coordinates}\label{ModesC}
The mode functions in Eq. (\ref{UmodeR0x}) can be expressed in terms of $R_{1}$ coordinates. Using $ x^{\prime}=x-\ell$, we have the two Rindler frames connected as
\begin{equation}\label{Reln}
\begin{aligned}
\frac{1}{a_0} e^{a_0 \xi_0} \sinh \left(a_0 \eta_0\right) & =\frac{1}{a_1} e^{a_1 \xi_1} \sinh \left(a_1 \eta_1\right)\,, \\
\frac{1}{a_0} e^{a_0 \xi_0} \cosh \left(a_0 \eta_0\right) & =\frac{1}{a_1} e^{a_1 \xi_1} \cosh \left(a_1 \eta_1\right)+\ell\,.
\end{aligned}
\end{equation}
Using these relations one obtains (for more details see \cite{Lochan:2025mru})
\begin{equation}\label{eEta}
e^{i\omega_{0}\eta_{0}}=\left(\frac{e^{a_1 \eta_1}+g_{1}}{e^{-a_1 \eta_1}+g_{1}}\right)^{\frac{i\omega_{0}}{2a_{0}}}\,,
\end{equation}
and
\begin{equation}\label{eXi}
e^{i\omega_{0}\xi_{0}}=\left( \frac{a_{0}}{a_{1}}\right)^{\frac{i\omega_{0}}{a_{0}}}
e^{\frac{i\omega_{0}}{a_{0}}a_{1}\xi_{1}}[(g_{1}+e^{-a_{1}\eta_{1}})(g_{1}+e^{a_{1}\eta_{1}})]^{\frac{i\omega_{0}}{2a_{0}}}\,.
\end{equation}
Finally use of Eq. (\ref{eEta}) and (\ref{eXi}) leads to Eq. (\ref{UmodeR0}).

\section{Calculation of entangling term}\label{CalcR0}
\subsection{Evaluation of Eq. (\ref{EE1x})}

In this Appendix, we calculate the entangling term $\mathcal{E}$ in Eq. (\ref{EE0}). One can use the Wightman propagator in Eq. (\ref{GR0}) to evaluate the entangling term. We obtain
\begin{equation}\label{ }\begin{aligned}
\mathcal{E}&=
-\int\frac{d\omega_{0}}{4\pi\omega_{0}}
\int_{-\infty}^{\infty}d\tau_{1}\int_{\tau_{1}}^{\infty}d\tau_{0}\,e^{i\Omega(\tau_{0}+\tau_{1})}
e^{-i\omega_{0}\tau_{0}}
\left( \frac{a_{0}}{a_{1}}\right)^{-\frac{i\omega_{0}}{a_{0}}}e^{\frac{i\omega_{0}}{a_{0}}a_{1}\tau_{1}}(1+a_{1}\ell e^{a_{1}\tau_{1}})^{-\frac{i\omega_{0}}{a_{0}}}\\&~~~~
-
\int\frac{d\omega_{0}}{4\pi\omega_{0}}
\int_{-\infty}^{\infty}d\tau_{1}\int_{\tau_{1}}^{\infty}d\tau_{0}\,e^{i\Omega(\tau_{0}+\tau_{1})}
e^{-i\omega_{0}\tau_{0}}
	\left( \frac{a_{0}}{a_{1}}\right)^{\frac{i\omega_{0}}{a_{0}}} e^{\frac{i\omega_{0}}{a_{0}}a_{1}\tau_{1}}(1+a_{1}\ell e^{-a_{1}\tau_{1}})^{\frac{i\omega_{0}}{a_{0}}}\\&~~~~
	-	
	\int\frac{d\omega_{0}}{4\pi\omega_{0}}
\int_{-\infty}^{\infty}d\tau_{1}\int_{-\infty}^{\tau_{1}}d\tau_{0}\,e^{i\Omega(\tau_{0}+\tau_{1})}e^{i\omega_{0}\tau_{0}}
\left( \frac{a_{0}}{a_{1}}\right)^{\frac{i\omega_{0}}{a_{0}}}e^{-\frac{i\omega_{0}}{a_{0}}a_{1}\tau_{1}}(1+a_{1}\ell e^{a_{1}\tau_{1}})^{\frac{i\omega_{0}}{a_{0}}}\\&~~~~
-	
	\int\frac{d\omega_{0}}{4\pi\omega_{0}}
\int_{-\infty}^{\infty}d\tau_{1}\int_{-\infty}^{\tau_{1}}d\tau_{0}\,e^{i\Omega(\tau_{0}+\tau_{1})}e^{i\omega_{0}\tau_{0}}
	\left( \frac{a_{0}}{a_{1}}\right)^{-\frac{i\omega_{0}}{a_{0}}} e^{-\frac{i\omega_{0}}{a_{0}}a_{1}\tau_{1}}(1+a_{1}\ell e^{-a_{1}\tau_{1}})^{-\frac{i\omega_{0}}{a_{0}}}\\& \equiv -A-B-C-D\,.
\end{aligned}\end{equation}
Now we will evaluate the terms $A,~B,~C$ and $D$ one by one. For all terms, we first perform $\tau_{0}$-integral, then $\tau_{1}$-integral to obtain the expression in Eq. (\ref{EE1x}). To perform $\tau_{0}$ integral, we will introduce an infinitesimal  parameter $\epsilon$  to converge the integration. Later we will set the limit $\epsilon\to0$. After that, to perform $\tau_{1}$ integration, we will introduce another two infinitesimal parameters $\nu$ and $\lambda$ with conditions $\nu>\lambda$. Again we will take the limits $\nu\to0$ and $\lambda\to0$. The similar idea was adopted in \cite{Lochan:2025mru,PhysRevD.111.045023} as well.

Let us evaluate first the quantity $A$:
\begin{equation}
A=\int_{0}^{\infty}\frac{d\omega_{0}}{4\pi\omega_{0}}\left( \frac{a_{0}}{a_{1}}\right)^{-\frac{i\omega_{0}}{a_{0}}}
\int_{-\infty}^{\infty}d\tau_{1}
e^{i\Omega\tau_{1}+\frac{i\omega_{0}}{a_{0}}a_{1}\tau_{1}}(1+a_{1}\ell e^{a_{1}\tau_{1}})^{-\frac{i\omega_{0}}{a_{0}}}
\int_{\tau_{1}}^{\infty}d\tau_{0}\,e^{i\tau_{0}(\Omega-\omega_{0}+i\epsilon)}\,.
\end{equation}
Performing $\tau_{0}$-integral we obtain
\begin{equation}\label{EntA00}
A=\int_{0}^{\infty}\frac{d\omega_{0}}{4\pi\omega_{0}}\frac{i\left( \frac{a_{0}}{a_{1}}\right)^{-\frac{i\omega_{0}}{a_{0}}}}{a_{1}(\Omega-\omega_{0}+i\epsilon)}
\int_{-\infty}^{\infty}d(a_{1}\tau_{1})
e^{(\lambda+i(\frac{2\Omega-\omega_{0}}{a_{1}}+\frac{\omega_{0}}{a_{0}}))a_{1}\tau_{1}}(1+a_{1}\ell e^{a_{1}\tau_{1}})^{-(\nu+\frac{i\omega_{0}}{a_{0}})}\,.
\end{equation}
Here we introduce $\nu\to0$ and $\lambda\to0$ with $\nu>\lambda$. We will utilize the following formula to evaluate this integral  \cite{TablesFT}.
\begin{equation}\label{FTF1x}\begin{aligned}
\int_{-\infty}^{\infty}dx\,e^{-(\lambda-iy)x}(1+Ae^{-x})^{-\nu}
&=\int_{-\infty}^{\infty}dx\,e^{(\lambda-iy)x}(1+Ae^{x})^{-\nu}\\&=A^{-(\lambda-iy)}B[\lambda-iy,\nu-\lambda+iy]~~~~~~~~~~~~[\text{Re}(\nu)>\text{Re}(\lambda)]\,.
\end{aligned}\end{equation}
Thus Eq. (\ref{EntA00}) becomes 
\begin{equation}\label{EntA}
A=\int_{0}^{\infty}\frac{d\omega_{0}}{4\pi\omega_{0}}\frac{i\left( \frac{a_{0}}{a_{1}}\right)^{-\frac{i\omega_{0}}{a_{0}}}(a_{1}\ell)^{-i(\frac{2\Omega-\omega_{0}}{a_{1}}+\frac{\omega_{0}}{a_{0}})}B[i(\frac{2\Omega-\omega_{0}}{a_{1}}+\frac{\omega_{0}}{a_{0}}),\nu-i(\frac{2\Omega-\omega_{0}}{a_{1}})]}{a_{1}(\Omega-\omega_{0}+i\epsilon)}\,.
\end{equation}
Similarly, we evaluate the quantity $B$:
\begin{equation}\label{ }
B=\int_{0}^{\infty}\frac{d\omega_{0}}{4\pi\omega_{0}}\left( \frac{a_{0}}{a_{1}}\right)^{\frac{i\omega_{0}}{a_{0}}}
\int_{-\infty}^{\infty}d\tau_{1}e^{i(\Omega+\frac{\omega_{0}}{a_{0}}a_{1})\tau_{1}}
(1+a_{1}\ell e^{-a_{1}\tau_{1}})^{\frac{i\omega_{0}}{a_{0}}}
\int_{\tau_{1}}^{\infty}	d\tau_{0}\,e^{i(\Omega-\omega_{0}+i\epsilon)\tau_{0}}~,
\end{equation}
where, we have $\epsilon\to0$ at the end. Performing $\tau_{0}$-integral we obtain
\begin{equation}\label{ }
B=\int\frac{d\omega_{0}}{4\pi\omega_{0}}
	\frac{i\left( \frac{a_{0}}{a_{1}}\right)^{\frac{i\omega_{0}}{a_{0}}}}{(\Omega-\omega_{0}+i\epsilon)}
\int_{-\infty}^{\infty}d\tau_{1}\,e^{-(\lambda-i(\frac{2\Omega-\omega_{0}}{a_{1}}+\frac{\omega_{0}}{a_{0}}))a_{1}\tau_{1}}(1+a_{1}\ell e^{-a_{1}\tau_{1}})^{-(\nu-\frac{i\omega_{0}}{a_{0}})}~.
\end{equation}
Then using Eq. (\ref{FTF1x}) one can perform $\tau_{1}$ integration. This yields 
\begin{equation}\label{EntB}
B=\int_{0}^{\infty}\frac{d\omega_{0}}{4\pi\omega_{0}}
	\frac{i\left( \frac{a_{0}}{a_{1}}\right)^{\frac{i\omega_{0}}{a_{0}}}	(a_{1}\ell)^{+i(\frac{2\Omega-\omega_{0}}{a_{1}}+\frac{\omega_{0}}{a_{0}})}
	B[-i(\frac{2\Omega-\omega_{0}}{a_{1}}+\frac{\omega_{0}}{a_{0}}),\nu
	+i(\frac{2\Omega-\omega_{0}}{a_{1}})]}{a_{1}(\Omega-\omega_{0}+i\epsilon)}\,.
\end{equation}
Then we evaluate the quantity $C$:
\begin{equation}\label{ }
C=\int_{0}^{\infty}\frac{d\omega_{0}}{4\pi\omega_{0}}\left( \frac{a_{0}}{a_{1}}\right)^{\frac{i\omega_{0}}{a_{0}}}
\int_{-\infty}^{\infty}d\tau_{1}
e^{i\Omega\tau_{1}}
e^{-\frac{i\omega_{0}}{a_{0}}a_{1}\tau_{1}}(1+a_{1}\ell e^{a_{1}\tau_{1}})^{\frac{i\omega_{0}}{a_{0}}}
\int_{-\infty}^{\tau_{1}}d\tau_{0}\,
e^{i(\Omega+\omega_{0}-i\epsilon)\tau_{0}}~.
\end{equation}
Performing $\tau_{0}$ integral one finds
\begin{equation}\label{ }
C=\int_{0}^{\infty}\frac{d\omega_{0}}{4\pi\omega_{0}}
\frac{(-i)\left( \frac{a_{0}}{a_{1}}\right)^{\frac{i\omega_{0}}{a_{0}}}}{(\Omega+\omega_{0}-i\epsilon)}\int_{-\infty}^{\infty}d\tau_{1}e^{(\lambda+i(\frac{2\Omega+\omega_{0}}{a_{1}}-\frac{\omega_{0}}{a_{0}}))a_{1}\tau_{1}}(1+a_{1}\ell e^{a_{1}\tau_{1}})^{-(\nu-\frac{i\omega_{0}}{a_{0}})}~.
\end{equation}
Then performing $\tau_{1}$ integral using Eq. (\ref{FTF1x}), we obtain
\begin{equation}\label{EntC00}\begin{aligned}
C&=\int_{0}^{\infty}\frac{d\omega_{0}}{4\pi\omega_{0}}
\frac{(-i)\left( \frac{a_{0}}{a_{1}}\right)^{\frac{i\omega_{0}}{a_{0}}}(a_{1}\ell)^{-i(\frac{2\Omega+\omega_{0}}{a_{1}}-\frac{\omega_{0}}{a_{0}})}
B[i(\frac{2\Omega+\omega_{0}}{a_{1}}-\frac{\omega_{0}}{a_{0}}),\,\nu-i\frac{2\Omega+\omega_{0}}{a_{1}}]}{a_{1}(\Omega+\omega_{0}-i\epsilon)}\,.
\end{aligned}\end{equation}
Here we can write $\frac{1}{(\Omega+\omega_{0}-i\epsilon)}=P\frac{1}{\Omega+\omega_{0}}+i\pi\delta(\Omega+\omega_{0})$. Since both $\Omega$ and $\omega_{0}$ are positive, thus $\delta(\Omega+\omega_{0})=0$. Again we can express $P\frac{1}{\Omega+\omega_{0}}=\frac{1}{(\Omega+\omega_{0}+i\epsilon')}+i\pi\delta(\Omega+\omega_{0})$. Thus we can write Eq. (\ref{EntC00}) as
\begin{equation}\label{EntC}\begin{aligned}
C&=\int_{0}^{\infty}\frac{d\omega_{0}}{4\pi\omega_{0}}
\frac{(-i)\left( \frac{a_{0}}{a_{1}}\right)^{\frac{i\omega_{0}}{a_{0}}}(a_{1}\ell)^{-i(\frac{2\Omega+\omega_{0}}{a_{1}}-\frac{\omega_{0}}{a_{0}})}
B[i(\frac{2\Omega+\omega_{0}}{a_{1}}-\frac{\omega_{0}}{a_{0}}),\,\nu-i\frac{2\Omega+\omega_{0}}{a_{1}}]}{a_{1}(\Omega+\omega_{0}+i\epsilon')}\\&=
\int_{-\infty}^{0}\frac{d\omega_{0}}{4\pi\omega_{0}}
\frac{i\left( \frac{a_{0}}{a_{1}}\right)^{-\frac{i\omega_{0}}{a_{0}}}(a_{1}\ell)^{-i(\frac{2\Omega-\omega_{0}}{a_{1}}+\frac{\omega_{0}}{a_{0}})}
B[i(\frac{2\Omega-\omega_{0}}{a_{1}}+\frac{\omega_{0}}{a_{0}}),\,\nu-i\frac{2\Omega-\omega_{0}}{a_{1}}]}{a_{1}(\Omega-\omega_{0}+i\epsilon')}\,.
\end{aligned}\end{equation}
Here, in the last equality, we used variable change $\omega_{0}\to-\omega_{0}$. Also note that the $\epsilon'$ in Eq. (\ref{EntC}) and $\epsilon$ in Eq. (\ref{EntA}) may be different in general. However in both term we have $\epsilon,\epsilon' \to0$. Hence we can take equal $\epsilon'=\epsilon$ and add Eq.(\ref{EntC}) and (\ref{EntA}). Then we obtain
\begin{equation}\label{C1nt}\begin{aligned}
A+C&=
\int_{-\infty}^{\infty}\frac{d\omega_{0}}{4\pi\omega_{0}}
\frac{i\left( \frac{a_{0}}{a_{1}}\right)^{-\frac{i\omega_{0}}{a_{0}}}(a_{1}\ell)^{-i(\frac{2\Omega-\omega_{0}}{a_{1}}+\frac{\omega_{0}}{a_{0}})}B[i(\frac{2\Omega-\omega_{0}}{a_{1}}+\frac{\omega_{0}}{a_{0}}),\nu
-i(\frac{2\Omega-\omega_{0}}{a_{1}})]}{a_{1}(\Omega-\omega_{0}+i\epsilon)}\,.
\end{aligned}\end{equation}
Finally, we evaluate the quantity $D$:
\begin{equation}\label{ }
D=\int_{0}^{\infty}\frac{d\omega_{0}}{4\pi\omega_{0}}	\left( \frac{a_{0}}{a_{1}}\right)^{-\frac{i\omega_{0}}{a_{0}}}
\int_{-\infty}^{\infty}d\tau_{1}e^{i\Omega\tau_{1}}
 e^{-\frac{i\omega_{0}}{a_{0}}a_{1}\tau_{1}}(1+a_{1}\ell e^{-a_{1}\tau_{1}})^{-\frac{i\omega_{0}}{a_{0}}}
 \int_{-\infty}^{\tau_{1}}d\tau_{0}\,e^{i(\Omega+\omega_{0}-i\epsilon)\tau_{0}}~.
\end{equation}
Performing $\tau_{0}$ integration, we obtain
\begin{equation}\label{ }
D=\int_{0}^{\infty}\frac{d\omega_{0}}{4\pi\omega_{0}}\frac{(-i)\left( \frac{a_{0}}{a_{1}}\right)^{-\frac{i\omega_{0}}{a_{0}}}}{(\Omega+\omega_{0}-i\epsilon)}\int_{-\infty}^{\infty}d\tau_{1}
	e^{-(\lambda-i(\frac{2\Omega+\omega_{0}}{a_{1}}-\frac{\omega_{0}}{a_{0}}))a_{1}\tau_{1}}(1+a_{1}\ell e^{-a_{1}\tau_{1}})^{-(\nu+\frac{i\omega_{0}}{a_{0}})}~.
\end{equation}
Then performing $\tau_{1}$ integral using Eq. (\ref{FTF1x}), we obtain
\begin{equation}\label{ }\begin{aligned}
D&=\int_{0}^{\infty}\frac{d\omega_{0}}{4\pi\omega_{0}}\frac{(-i)\left( \frac{a_{0}}{a_{1}}\right)^{-\frac{i\omega_{0}}{a_{0}}}(a_{1}\ell)^{+i(\frac{2\Omega+\omega_{0}}{a_{1}}-\frac{\omega_{0}}{a_{0}})}
B[-i(\frac{2\Omega+\omega_{0}}{a_{1}}-\frac{\omega_{0}}{a_{0}}),\nu+i(\frac{2\Omega+\omega_{0}}{a_{1}})]}{a_{1}(\Omega+\omega_{0}-i\epsilon)}\\&=
\int_{-\infty}^{0}\frac{d\omega_{0}}{4\pi\omega_{0}}\frac{i\left( \frac{a_{0}}{a_{1}}\right)^{\frac{i\omega_{0}}{a_{0}}}(a_{1}\ell)^{+i(\frac{2\Omega-\omega_{0}}{a_{1}}+\frac{\omega_{0}}{a_{0}})}
B[-i(\frac{2\Omega-\omega_{0}}{a_{1}}+\frac{\omega_{0}}{a_{0}}),\nu+i(\frac{2\Omega-\omega_{0}}{a_{1}})]}{a_{1}(\Omega-\omega_{0}-i\epsilon)}~.
\end{aligned}\end{equation}
Here again, we can use $\frac{1}{\Omega-\omega_{0}-i\epsilon}=P\frac{1}{\Omega-\omega_{0}}+i\pi\delta(\Omega-\omega_{0})$ with the information $\Omega>0$ and $\omega_{0}<0$. Thus we have $\delta(\Omega-\omega_{0})=0$. Then again we can express the principle value as $P\frac{1}{\Omega-\omega_{0}}=\frac{1}{\Omega-\omega_{0}+i\epsilon'}+i\pi\delta(\Omega-\omega_{0})$. Thus we obtain
\begin{equation}\label{EntD}\begin{aligned}
D&=
\int_{-\infty}^{0}\frac{d\omega_{0}}{4\pi\omega_{0}}\frac{i\left( \frac{a_{0}}{a_{1}}\right)^{\frac{i\omega_{0}}{a_{0}}}(a_{1}\ell)^{+i(\frac{2\Omega-\omega_{0}}{a_{1}}+\frac{\omega_{0}}{a_{0}})}
B[-i(\frac{2\Omega-\omega_{0}}{a_{1}}+\frac{\omega_{0}}{a_{0}}),\nu+i(\frac{2\Omega-\omega_{0}}{a_{1}})]}{a_{1}(\Omega-\omega_{0}+i\epsilon')}\,.
\end{aligned}\end{equation}
Now Adding Eq.(\ref{EntB}) and (\ref{EntD}), we obtain
\begin{equation}\label{C2nt}\begin{aligned}
B+D&=\int_{-\infty}^{\infty}\frac{d\omega_{0}}{4\pi\omega_{0}}\frac{i\left( \frac{a_{0}}{a_{1}}\right)^{\frac{i\omega_{0}}{a_{0}}}(a_{1}\ell)^{+i(\frac{2\Omega-\omega_{0}}{a_{1}}+\frac{\omega_{0}}{a_{0}})}
B[-i(\frac{2\Omega-\omega_{0}}{a_{1}}+\frac{\omega_{0}}{a_{0}}),\nu+i(\frac{2\Omega-\omega_{0}}{a_{1}})]}{a_{1}(\Omega-\omega_{0}+i\epsilon)}\,.
\end{aligned}\end{equation}
Therefore adding Eq. (\ref{C1nt}) and (\ref{C2nt}) with $a_{0}=a_{1}$, we obtain Eq. (\ref{EE1x}).

\subsection{Evaluation of contour integrals in Eq. (\ref{EE1x})}
Let us denote $I_{AC}$ and $I_{BD}$ as the integrands of the integrals in Eq. (\ref{EE1x}), respectively:
\begin{equation}\label{EE1xINT}\begin{aligned}
I_{AC}&=\frac{i}{4\pi\omega_{0}}
\frac{(a_{1}\ell)^{-i(\frac{2\Omega}{a_{1}})}B[i(\frac{2\Omega}{a_{1}}),\nu
-i(\frac{2\Omega-\omega_{0}}{a_{1}})]}{a_{1}(\Omega-\omega_{0}+i\epsilon)}\,;
\\I_{BD}&=
\frac{i}{4\pi\omega_{0}}\frac{(a_{1}\ell)^{+i(\frac{2\Omega}{a_{1}})}
B[-i(\frac{2\Omega}{a_{1}}),\nu+i(\frac{2\Omega-\omega_{0}}{a_{1}})]}{a_{1}(\Omega-\omega_{0}+i\epsilon)}\,.
\end{aligned}\end{equation}
We can express the beta functions in terms of gamma functions (as $B[z_{1},z_{2}]=\Gamma[z_{1}]\Gamma[z_{1}]/\Gamma[z_{1}+z_{2}]$), and rewrite the integrands as
\begin{equation}\label{EE1xINT1}\begin{aligned}
I_{AC}&=\frac{-1}{4\pi a_{1}}
\frac{(a_{1}\ell)^{-i(\frac{2\Omega}{a_{1}})}\Gamma[i(\frac{2\Omega}{a_{1}})]\Gamma[\nu
-i(\frac{2\Omega-\omega_{0}}{a_{1}})]}{a_{1}(\Omega-\omega_{0}+i\epsilon)\Gamma[1+i\frac{\omega_{0}}{a_{1}}]}\,;
\\I_{BD}&=
\frac{1}{4\pi a_{1}}\frac{(a_{1}\ell)^{+i(\frac{2\Omega}{a_{1}})}
\Gamma[-(\frac{2\Omega}{a_{1}})]\Gamma[\nu+i(\frac{2\Omega-\omega_{0}}{a_{1}})]}{a_{1}(\Omega-\omega_{0}+i\epsilon)\Gamma[1-i\frac{\omega_{0}}{a_{1}}]}\,.
\end{aligned}\end{equation}
Here we set the limit $\nu\to0$ in the denominator and used $z\,\Gamma[z]=\Gamma[1+z]$.
In the numerator, the quantity $\nu$ in the gamma function still kept there as it will help us to determine the position of poles in the complex $\omega_{0}$-plane. 

 It is easy to see that  both integrands $I_{AC}$ and $I_{BD}$ vanishes at $\omega_{0}\to\pm{i\infty}$. The poles for these quantities arise at $\omega_{0}=\Omega+i\epsilon$ and when the gamma functions in the numerator become divergent {\ie} when the gamma functions have argument $z=-n$ with $n\geq0$ (as $\Gamma[-n]$ is divergent for $n\geq0$).
 The poles of the integrands are given below.\\
Poles of $I_{AC}$:
\begin{eqnarray}
\omega_{0}=\Omega+i\epsilon;~~~\omega_{0}=2\Omega+ia_{1}\nu+ia_{1}n~~~(n\geq0)\,.
\end{eqnarray}
Poles of $I_{BD}$:
\begin{eqnarray}
\omega_{0}=\Omega+i\epsilon;~~~\omega_{0}=2\Omega-ia_{1}\nu-ia_{1}n~~~(n\geq0)\,.
\end{eqnarray}

\subsubsection{For Upper Complex $\omega_{0}$-plane}
 If we consider upper half of the complex $\omega_{0}$-plane, the following Residue contributions will be added.
 For the integrand $I_{BD}$ only pole in the upper complex $\omega_{0}$-plane is $\omega_{0}=\Omega+i\epsilon$. The corresponding residue contribution is 
\begin{equation}\label{BDup}
2\pi i \,\,\text{Res}(I_{BD})_{\omega_{0}=\Omega+i\epsilon}
=\frac{ (a_{1} l)^{\frac{2 i \Omega }{a_{1}}} \Gamma \left(\frac{i \Omega }{a_{1}}\right) \Gamma \left(-\frac{2 i \Omega }{a_{1}}\right)}{2a_{1}\Omega \Gamma \left(-\frac{i \Omega }{a_{1}}\right)}\,.
\end{equation}
However for $I_{AC}$, poles are at $\omega_{0}=\Omega+i\epsilon$ and $\omega_{0}=2\Omega+ia_{1}\nu+ia_{1}n~~(n\geq0)$. The residue contribution at $\omega_{0}=\Omega+i\epsilon$ is 
\begin{equation}\label{ACP1}
2\pi i \,\,\text{Res}(I_{AC})_{\omega_{0}=\Omega+i\epsilon}=\frac{(a_{1} l)^{-\frac{2 i \Omega }{a_{1}}} \Gamma \left(-\frac{i \Omega }{a_{1}}\right) \Gamma \left(\frac{2 i \Omega }{a_{1}}\right)}{2 a_{1}\Omega \Gamma \left(\frac{i \Omega }{a_{1}}\right)}\,.
\end{equation}
The residue contributions at $\omega_{0}=2\Omega+ia_{1}\nu+ia_{1}n~~(n\geq0)$ are given by
\begin{equation}\label{ }
2\pi i\,\, \text{Res}(I_{AC})_{\omega_{0}=\Omega+ia_{1}\nu+ia_{1}n}=
-\frac{i (-1)^n (a_{1} l)^{-\frac{2 i \Omega }{a_{1}}} \Gamma \left(\frac{2 i \Omega }{a_{1}}\right)}{2a_{1} n! (a_{1} n-i \Omega ) \Gamma \left(\frac{-n a_{1}+a_{1}+2 i \Omega }{a_{1}}\right)}~.
\end{equation}
Summing over all $n\geq0$, we obtain
\begin{equation}\label{ACPS1}\begin{aligned}
2\pi i \sum_{n=0}^{\infty}\text{Res}(I_{AC})_{\omega_{0}=\Omega+ia_{1}\nu+ia_{1}n}&=
\frac{2^{-1+\frac{2 i \Omega }{a_{1}}} (a_{1} l)^{-\frac{2 i \Omega }{a_{1}}} \Gamma \left(\frac{a_{1}-i \Omega }{a_{1}}\right) \Gamma \left(\frac{a_{1}+2 i \Omega }{2 a_{1}}\right) \Gamma \left(\frac{2 i \Omega }{a_{1}}\right)}{\sqrt{\pi } a_{1} \Omega  \Gamma \left(\frac{a_{1}+2 i \Omega }{a_{1}}\right)}\\
&=
\frac{-(a_{1} l)^{-\frac{2 i \Omega }{a_{1}}}   \Gamma \left(-\frac{i \Omega }{a_{1}}\right)  \Gamma \left(\frac{2 i \Omega }{a_{1}}\right)}{2a_{1}\Omega\Gamma \left(\frac{i \Omega }{a_{1}}\right)}\,.
\end{aligned}\end{equation}
In the above we used the following formula \cite{NIST:DLMF}
\begin{equation}\label{GF2z}
\Gamma[2z]=\frac{2^{2z-1}\Gamma[z]\Gamma[z+\frac{1}{2}]}{\sqrt{\pi}}\,.
\end{equation}

Thus contributions from Eq. (\ref{ACP1}) and  Eq. (\ref{ACPS1}) cancel out. So the first integral in Eq. (\ref{EE1x}) vanishes. Therefore, only second integral gives a non-vanishing contribution. Taking absolute value of this contribution in Eq. (\ref{BDup}), we obtain the entangling term in Eq. (\ref{EEF}).

\subsubsection{For Lower Complex $\omega_{0}$-plane}
If we consider lower half of the complex $\omega_{0}$-plane, only second integral will contribute in Eq. (\ref{EE1x}) as there is no poles exists in lower half for $I_{AC}$. Thus residue contribution for $\omega_{0}=\Omega-ia_{1}\nu-ia_{1}n~(n\geq0)$ is
 \begin{equation}\label{ }\begin{aligned}
-2\pi i \,\,\text{Res}(I_{BD})_{\omega_{0}=\Omega-ia_{1}\nu-ia_{1}n}=
-\frac{i (-1)^n (a_{1} l)^{\frac{2 i \Omega }{a_{1}}} \Gamma \left(-\frac{2 i \Omega }{a_{1}}\right)}{2 a_{1} n! (a_{1} n+i \Omega ) \Gamma \left(\frac{-n a_{1}+a_{1}-2 i \Omega }{a_{1}}\right)}\,.
\end{aligned}\end{equation}
Adding all the poles with $n\geq0$, we obtain
\begin{equation}\label{BDdown}\begin{aligned}
-2\pi i \sum_{n=0}^{\infty}\text{Res}(I_{BD})_{\omega_{0}=\Omega-ia_{1}\nu-ia_{1}n}
&=
-\frac{2^{-1-\frac{2 i \Omega }{a_{1}}} (a_{1} l)^{\frac{2 i \Omega }{a_{1}}} \Gamma \left(\frac{a_{1}+i \Omega }{a_{1}}\right) \Gamma \left(\frac{a_{1}-2 i \Omega }{2 a_{1}}\right) \Gamma \left(-\frac{2 i \Omega }{a_{1}}\right)}{\sqrt{\pi } a_{1}\Omega  \Gamma \left(\frac{a_{1}-2 i \Omega }{a_{1}}\right)}\,.
\end{aligned}\end{equation}
Simplifying the above expression using Eq. (\ref{GF2z}) and taking the absolute value, we again obtain Eq. (\ref{EEF}).
Thus, as expected, we see that results are same irrespective of choice of upper or lower plane.

\section{Alternative definition of mutual information and quantum discord}\label{MIQD}
The calculations presented here are already outlined in \cite{PhysRevA.77.042303, PhysRevA.81.042105}. Here we present the calculations specific to our density matrix given  in Eq. (\ref{eq:detector-density-matrix}).
The total classical correlation is defined as Max$[\mathcal{J}]$, where the quantity $\mathcal{J}$ is given in Eq. (\ref{DefJ}). One may write Max$[\mathcal{J}]$ as
\begin{equation}\label{DefJMax}\begin{aligned}
\text{Max}_{\{B_{k}\}}\mathcal{J}(\rho_{AB})=S(\rho_{A})-\text{Min}_{\{B_{k}\}}[S(\rho|\{B_{k}\})]\,,\end{aligned}\end{equation}
where $S(\rho|\{B_{k}\})$ is known as conditioned entropy and $\rho_{A}=\text{Tr}_{B}[\rho_{AB}]$. The eigenvalues of $\rho_{A}$ are \cite{Koga:2018the}:
\begin{equation}\label{ }\begin{aligned}
\lambda_{\pm}(\rho_{A})=\lambda^{2}\mathcal{P}_{A},~1-\lambda^{2}\mathcal{P}_{A}\,.\end{aligned}\end{equation}
Hence we obtain
\begin{equation}\label{SA}
S(\rho_{A})=-\lambda^{2}\mathcal{P}_{A}\log(\lambda^{2}\mathcal{P}_{A})-(1-\lambda^{2}\mathcal{P}_{A})\log(1-\lambda^{2}\mathcal{P}_{A})\,.
\end{equation}

To evaluate the conditioned entropy $S(\rho|\{B_{k}\})$, let us first define the local projective measurement for subsystem $B$ as $\Pi_k=|k\rangle\langle k|\,,$ with $k=0,\,1$ corresponding to states $|g_{B}\rangle$ and $|e_{B}\rangle$. Then any von-Neumann measurement for subsystem $B$ can be written as
\begin{equation}\label{ }
B_k=V \Pi_k V^{\dagger}\,;
\end{equation}
where $V$ is a unitary matrix in $\mathrm{U}(2)$, can be written as
\begin{equation}\label{ }
V=t I+i \vec{y} \cdot\vec{\sigma}\,,
\end{equation}
with $t \in \mathbb{R}, \vec{y}=\left(y_1, y_2, y_3\right) \in \mathbb{R}^3$, and 
\begin{equation}\label{kl}\begin{aligned}
(t^2+y_3^2)+(y_1^2+y_2^2)=k+l=1.
\end{aligned}\end{equation}
In the expression of the unitary matrix $V$, $I$ is the identity operator and $\sigma$'s are Pauli matrices. Here for later convenience, we defined 
\begin{equation}\label{klx}\begin{aligned}
 k=(t^2+y_3^2);~~~l=(y_1^2+y_2^2).
\end{aligned}
\end{equation}
After the measurement $\left\{B_k\right\}$, the state $\rho_{AB}$ will change to the ensemble $\left\{p_k,\,\rho_k\right\}$ with
\begin{equation}\label{ }
\rho_k:=\frac{1}{p_k}\left(I \otimes B_k\right) \rho_{AB}\left(I \otimes B_k\right)\,,
\end{equation}
and $p_k=\text{Tr}\left(I \otimes B_k\right) \rho_{AB}\left(I \otimes B_k\right)$. We need to evaluate $\rho_k$ and $p_k$. For this purpose, we write
\begin{equation}\label{pkrk}
\begin{aligned}
p_k \rho_k & =\left(I \otimes B_k\right) \rho_{AB}\left(I \otimes B_k\right)=\left[I \otimes\left(V \Pi_k V^{\dagger}\right)\right] \rho_{AB}\left[I \otimes\left(V \Pi_k V^{\dagger}\right)\right] \\
& =(I \otimes V)(I \otimes \Pi_kV^{\dagger}) \rho_{AB}(I \otimes V \Pi_k)\left(I \otimes V^{\dagger}\right)\,,\\
\end{aligned}
\end{equation}
and \begin{equation}\label{pk0}\begin{aligned}
p_{k}=\text{Tr}[\left(I \otimes B_k\right) \rho_{AB}\left(I \otimes B_k\right)]\,.
\end{aligned}\end{equation}

For evaluation of quantity in Eq. (\ref{pkrk}), it would be  much convenient if we can express the density matrix in Eq. (\ref{eq:detector-density-matrix}) in terms of bases made out of Pauli matrices \cite{PhysRevA.81.042105}
\begin{equation}\label{rhoX0}\begin{aligned}
\rho_{AB}&=\frac{1}{4}\begin{pmatrix}
1+a_{3}+b_{3}+c_{3}&0&0&c_{1}-c_{2}\\
0&1+a_{3}-b_{3}-c_{3}&c_{1}+c_{2}&0\\
0&c^{\star}+c_{2}^{\star}&1-a_{3}+b_{3}-c_{3}&0\\
c^{\star}-c_{2}^{\star}&0&0&1-a_{3}-b_{3}+c_{3}
\end{pmatrix}\,,
\end{aligned}\end{equation}
where the elements are given by
\begin{equation}\label{abccc}\begin{aligned}
a_{3}&=-(1-2\lambda^{2}\mathcal{P}_{A})\,,\\
b_{3}&=-(1-2\lambda^{2}\mathcal{P}_{B})\,,\\
c_{3}&=(1-2\lambda^{2}\mathcal{P}_{A}-2\lambda^{2}\mathcal{P}_{B})\,,\\
c_{1}&=2\lambda^{2}(\mathcal{P}_{AB}+\mathcal{E})\,,\\
c_{2}&=2\lambda^{2}(\mathcal{P}_{AB}-\mathcal{E})\,.
\end{aligned}\end{equation}
Here $a_{3},~b_{3},~c_{3}$ are real numbers and $c_{1},~c_{2}$ are complex numbers.
Using these elements, we can rewrite the density matrix (Eq. (\ref{rhoX0}) ) in terms of $\sigma$-matrix bases as
\begin{equation}\label{rABss}\begin{aligned}
\rho_{AB}&=\frac{1}{4}\Bigg(
1+a_{3}\sigma_{z}\otimes{1}+b_{3}1\otimes\sigma_{z}+c_{3}\sigma_{z}\otimes\sigma_{z}+\text{Re}(c_{1})\sigma_{x}\otimes\sigma_{x}-\text{Im}(c_{1})\sigma_{y}\otimes\sigma_{x}+\text{Re}(c_{2})\sigma_{y}\otimes\sigma_{y}+\text{Im}(c_{2})\sigma_{x}\otimes\sigma_{y}\Bigg)\,.
\end{aligned}
\end{equation}
Now to evaluate the quantity $(I \otimes \Pi_kV^{\dagger}) \rho_{AB}(I \otimes V \Pi_k)$ in Eq. (\ref{pkrk}), we will utilise the following relations provided in \cite{PhysRevA.77.042303}:
\begin{equation}\label{V-op}
\begin{aligned}
V^{\dagger} \sigma_x V= & \left(t^2+y_1^2-y_2^2-y_3^2\right) \sigma_x +2\left(t y_3+y_1 y_2\right) \sigma_y+\underbrace{2\left(-t y_2+y_1 y_3\right)}_{z_{1}} \sigma_z
\\
V^{\dagger} \sigma_y V= & 2\left(-t y_3+y_1 y_2\right) \sigma_x+\left(t^2+y_2^2-y_1^2-y_3^2\right) \sigma_y  +\underbrace{2\left(t y_1+y_2 y_3\right)}_{z_{2}}  \sigma_z,
\\
V^{\dagger} \sigma_z V= & 2\left(t y_2+y_1 y_3\right) \sigma_x+2\left(-t y_1+y_2 y_3\right) \sigma_y  +\underbrace{\left(t^2+y_3^2-y_1^2-y_2^2\right)}_{z_{3}=k-l}  \sigma_z,
\end{aligned}
\end{equation}
 and 
\begin{equation}\label{Pi-op}\begin{aligned}
\Pi_{0}\sigma_{z}\Pi_{0}&=\Pi_{0}\,,\\
\Pi_{1}\sigma_{z}\Pi_{1}&=-\Pi_{1}\,,\\
\Pi_{j}\sigma_{k}\Pi_{j}&=0~~~~~~~~~~~~~~~~~~~~(k=x,\,y;~~j=0,\,1)\,.
\end{aligned}\end{equation}
Combining the relations in Eq. (\ref{V-op}) and (\ref{Pi-op}), we obtain
\begin{equation}\label{Pi-V-op}\begin{aligned}
\Pi_{0}V^{\dagger} \sigma_l V\Pi_{0}&=z_{l}\Pi_{0}\,,\\
\Pi_{1}V^{\dagger} \sigma_l V\Pi_{1}&=-z_{l}\Pi_{1}~~~~~~~~~~~~~~~~~~~~~~~(l=x,\,y,\,z)\,.
\end{aligned}\end{equation}

Now using Eq. (\ref{Pi-V-op}) and (\ref{rABss}), we can express the quantity $(I \otimes \Pi_kV^{\dagger}) \rho(I \otimes V \Pi_k)$ from Eq. (\ref{pkrk}) as
\begin{equation}\label{ }\begin{aligned}
\frac{1}{4}\Bigg[
I+a_{3}\sigma_{z}+b_{3}I(\pm z_{3})+c_{3}\sigma_{z}(\pm z_{3})+\text{Re}(c_{1})\sigma_{x}(\pm z_{1})-\text{Im}(c_{1})\sigma_{y}(\pm z_{1})+\text{Re}(c_{2})\sigma_{y}(\pm z_{2})+\text{Im}(c_{2})\sigma_{x}(\pm z_{2})
\Bigg]\otimes \Pi_k\,.
\end{aligned}\end{equation}
Here $\pm$ signs are for $k=0,1$, respectively. Thus Eq. (\ref{pkrk}) becomes
\begin{equation}\label{pkrk2}\begin{aligned}
p_{k}\rho_{k}&=
\frac{1}{4}\Bigg[
I+a_{3}\sigma_{z}+b_{3}I(\pm z_{3})+c_{3}\sigma_{z}(\pm z_{3})+\text{Re}(c_{1})\sigma_{x}(\pm z_{1})-\text{Im}(c_{1})\sigma_{y}(\pm z_{1})+\text{Re}(c_{2})\sigma_{y}(\pm z_{2})+\text{Im}(c_{2})\sigma_{x}(\pm z_{2})
\Bigg]\otimes V\Pi_kV^{\dagger}\\
&=\rho^{A}_{k}\otimes V\Pi_kV^{\dagger}\,.
\end{aligned}\end{equation}
Next using Eq. (\ref{pk0}), we obtain
\begin{equation}\label{pk1}\begin{aligned}
p_{k}&=\frac{1}{2}(1\pm b_{3}z_{3})\,.
\end{aligned}
\end{equation}
Therefore the matrices $\rho^{A}_{0,1}$ in Eq. (\ref{pkrk2}) can be expressed as
\begin{equation}\label{ }\begin{aligned}
\rho_{0}^{A}&=\frac{1}{2}\begin{pmatrix}
1+\frac{a_{3}+c_{3}z_{3}}{1+b_{3}z_{3}}
&
\frac{z_{1}\text{Re}(c_{1})+iz_{1}\text{Im}(c_{1})-iz_{2}\text{Re}(c_{2})+z_{2}\text{Im}(c_{2})}{1+b_{3}z_{3}}
\\\frac{z_{1}\text{Re}(c_{1})-iz_{1}\text{Im}(c_{1})+iz_{2}\text{Re}(c_{2})+z_{2}\text{Im}(c_{2})}{1+b_{3}z_{3}}
&1-\frac{a_{3}+c_{3}z_{3}}{1+b_{3}z_{3}}
\end{pmatrix}\\
&=\frac{1}{2}\begin{pmatrix}
1+\frac{a_{3}+c_{3}z_{3}}{1+b_{3}z_{3}}&
\frac{z_{1}c_{1}-iz_{2}c_{2}}{1+b_{3}z_{3}}\\
\frac{z_{1}c_{1}^{\star}+iz_{2}c_{2}^{\star}}{1+b_{3}z_{3}}&1-\frac{a_{3}+c_{3}z_{3}}{1+b_{3}z_{3}}
\end{pmatrix}\,;
\\
\rho_{1}^{A}&=\frac{1}{2}\begin{pmatrix}
1+\frac{a_{3}-c_{3}z_{3}}{1-b_{3}z_{3}}
&
\frac{-z_{1}\text{Re}(c_{1})-iz_{1}\text{Im}(c_{1})+iz_{2}\text{Re}(c_{2})-z_{2}\text{Im}(c_{2})}{1-b_{3}z_{3}}
\\\frac{-z_{1}\text{Re}(c_{1})+iz_{1}\text{Im}(c_{1})-iz_{2}\text{Re}(c_{2})-z_{2}\text{Im}(c_{2})}{1-b_{3}z_{3}}
&1-\frac{a_{3}-c_{3}z_{3}}{1-b_{3}z_{3}}
\end{pmatrix}\\
&=\frac{1}{2}\begin{pmatrix}
1+\frac{a_{3}-c_{3}z_{3}}{1-b_{3}z_{3}}
&
-\frac{z_{1}c_{1}-iz_{2}c_{2}}{1-b_{3}z_{3}}
\\-\frac{z_{1}c_{1}^{\star}+iz_{2}c_{2}^{\star}}{1-b_{3}z_{3}}
&1-\frac{a_{3}-c_{3}z_{3}}{1-b_{3}z_{3}}
\end{pmatrix}\,.\\
\end{aligned}\end{equation}

For a matrix in a form $\frac{1}{2} \begin{pmatrix}
 1+Y & X \\
 X^{\star} & 1-Y \\
\end{pmatrix}$, eigenvalues are known as $\frac{1}{2} \left(1\pm\sqrt{|X|^{2}+Y^2}\right)$.
Therefore, the eigenvalues of $\rho^{A}_{0,1}$ are obtained as
\begin{equation}\label{LR01A}\begin{aligned}
\lambda_{\pm}(\rho_{0}^{A})&=\frac{1}{2}\Bigg(1\pm
\frac{\sqrt{\left({a_{3}+c_{3}z_{3}}\right)^{2}+\left|{z_{1}c_{1}-iz_{2}c_{2}}\right|^{2}}}{1+b_{3}z_{3}}\Bigg)\,;\\
\lambda_{\pm}(\rho_{1}^{A})&=\frac{1}{2}\Bigg(1\pm
\frac{\sqrt{\left({a_{3}-c_{3}z_{3}}\right)^{2}+\left|-{z_{1}c_{1}-iz_{2}c_{2}}\right|^{2}}}{1-b_{3}z_{3}}\Bigg)\,.\\
\end{aligned}\end{equation}

Now remembering that $k+l=1$ and $k-l=z_{3}$ (see Eq. (\ref{kl}) and Eq. (\ref{V-op})), we can write 
\begin{equation}\label{kl2}\begin{aligned}
k=\frac{1+z_{3}}{2}\,,~~~~l=\frac{1-z_{3}}{2}\,.
\end{aligned}\end{equation}

Therefore, using Eq. (\ref{kl2}), (\ref{kl}) and (\ref{abccc}) we can write the probabilities in Eq. (\ref{pk1}) as 
\begin{equation}\label{pk2}\begin{aligned}
p_{0}&=\frac{1+b_{3}z_{3}}{2}
=\frac{(k+l)+(-1+2\lambda^{2}\mathcal{P}_{B})(k-l)}{2}
\\&=\lambda^{2}\mathcal{P}_{B}k+(1-\lambda^{2}\mathcal{P}_{B})l\,;
\\
p_{1}&=\frac{1-b_{3}z_{3}}{2}
=\frac{(k+l)-(-1+2\lambda^{2}\mathcal{P}_{B})(k-l)}{2}
\\&=
(1-\lambda^{2}\mathcal{P}_{B})k+\lambda^{2}\mathcal{P}_{B}l\,.
\end{aligned}\end{equation}
Also we can check
\begin{equation}\label{abcPPPX}\begin{aligned}
a_{3}+c_{3}z_{3}&=-(1-2\lambda^{2}\mathcal{P}_{A})(k+l)+(1-2\lambda^{2}\mathcal{P}_{A}-2\lambda^{2}\mathcal{P}_{B})(k-l)\\
&=-2[\lambda^{2}\mathcal{P}_{B}k+(1-2\lambda^{2}\mathcal{P}_{A}-\lambda^{2}\mathcal{P}_{B})l]\,;\\
a_{3}-c_{3}z_{3}&=-(1-2\lambda^{2}\mathcal{P}_{A})(k+l)-(1-2\lambda^{2}\mathcal{P}_{A}-2\lambda^{2}\mathcal{P}_{B})(k-l)\\
&=-2[\lambda^{2}\mathcal{P}_{B}l+(1-2\lambda^{2}\mathcal{P}_{A}-\lambda^{2}\mathcal{P}_{B})k]\,;\\
z_{1}c_{1}-iz_{2}c_{2}&=2\lambda^{2}\mathcal{P}_{AB}(z_{1}-iz_{2})+2\lambda^{2}\mathcal{E}(z_{1}+iz_{2})=2\lambda^{2}X\,.
\end{aligned}
\end{equation}
Finally, using  Eq. (\ref{pk2}) and (\ref{abcPPPX}), we can express the eigenvalues in Eq. (\ref{LR01A}) as
\begin{equation}\label{LR01A2}\begin{aligned}
\lambda_{\pm}(\rho_{0}^{A})&=\frac{1}{2}\Bigg(1\pm
\frac{\sqrt{[\lambda^{2}\mathcal{P}_{B}k+(1-2\lambda^{2}\mathcal{P}_{A}-\lambda^{2}\mathcal{P}_{B})l]^{2}+\lambda^{4}\left|X\right|^{2}}}{\lambda^{2}\mathcal{P}_{B}k+(1-\lambda^{2}\mathcal{P}_{B})l}\Bigg)\,;\\
\lambda_{\pm}(\rho_{1}^{A})&=\frac{1}{2}\Bigg(1\pm
\frac{\sqrt{[\lambda^{2}\mathcal{P}_{B}l+(1-2\lambda^{2}\mathcal{P}_{A}-\lambda^{2}\mathcal{P}_{B})k]^{2}+\lambda^{4}\left|X\right|^{2}}}{\lambda^{2}\mathcal{P}_{B}l+(1-\lambda^{2}\mathcal{P}_{B})k}\Bigg)\,.
\end{aligned}\end{equation}

Since in our calculation, we are only considering perturbative order up to $\lambda^{2}$, we expand these eigenvalues up to $\mathcal{O}(\lambda^{2})$. Then one obtains 
\begin{equation}\label{LR01A3}\begin{aligned}
\lambda_{\pm}(\rho_{0}^{A})&=\lambda^{2}\mathcal{P}_{A},~~1-\lambda^{2}\mathcal{P}_{A};\\
\lambda_{\pm}(\rho_{1}^{A})&=\lambda^{2}\mathcal{P}_{A},~~1-\lambda^{2}\mathcal{P}_{A}\,.
\end{aligned}
\end{equation}
Thus the conditional entropy becomes
\begin{equation}\label{SAcon}\begin{aligned}
S(\rho|\{B_{k}\})&=p_{0}S(\rho_{0}^{A})+p_{1}S(\rho_{1}^{A})=(p_{0}+p_{1})S(\rho_{0}^{A})
\\&=-\lambda^{2}\mathcal{P}_{A}\log(\lambda^{2}\mathcal{P}_{A})-(1-\lambda^{2}\mathcal{P}_{A})\log(1-\lambda^{2}\mathcal{P}_{A})\,.
\end{aligned}\end{equation}
It is independent of the parameters in the unitary matrix $V$ and independent of the choices of measurement basis $\{B_{k}\}$ up to $\mathcal{O}(\lambda^{2})$. Hence no minimization is required for the conditional entropy.

Note that for our analysis upto $\mathcal{O}(\lambda^2)$ the above expressions are independent of $\mathcal{P}_A$, $\mathcal{P}_{AB}$ and $\mathcal{E}$.
Thus from Eq. (\ref{DefJMax}), (\ref{SA}) and (\ref{SAcon}), one can see that the alternative definition of mutual information or the classical correlations vanishes up to $\mathcal{O}(\lambda^{2})$. So for our density matrix in Eq. (\ref{eq:detector-density-matrix}), quantum discord is equal to the quantum mutual information for perturbative analysis (in agreement with \cite{Lin:2024roh}).

\end{appendix}
\end{widetext}

\bibliographystyle{apsrev}

\bibliography{bibtexfile}

\begin{thebibliography}{80}
\expandafter\ifx\csname natexlab\endcsname\relax\def\natexlab#1{#1}\fi
\expandafter\ifx\csname bibnamefont\endcsname\relax
  \def\bibnamefont#1{#1}\fi
\expandafter\ifx\csname bibfnamefont\endcsname\relax
  \def\bibfnamefont#1{#1}\fi
\expandafter\ifx\csname citenamefont\endcsname\relax
  \def\citenamefont#1{#1}\fi
\expandafter\ifx\csname url\endcsname\relax
  \def\url#1{\texttt{#1}}\fi
\expandafter\ifx\csname urlprefix\endcsname\relax\def\urlprefix{URL }\fi
\providecommand{\bibinfo}[2]{#2}
\providecommand{\eprint}[2][]{\url{#2}}

\bibitem[{\citenamefont{Hotta}(2008)}]{Hotta:2008uk}
\bibinfo{author}{\bibfnamefont{M.}~\bibnamefont{Hotta}},
  \bibinfo{journal}{Phys. Rev. D} \textbf{\bibinfo{volume}{78}},
  \bibinfo{pages}{045006} (\bibinfo{year}{2008}), \eprint{arXiv:0803.2272}.

\bibitem[{\citenamefont{Hotta}(2009)}]{Hotta:2009}
\bibinfo{author}{\bibfnamefont{M.}~\bibnamefont{Hotta}},
  \bibinfo{journal}{Journal of the Physical Society of Japan}
  \textbf{\bibinfo{volume}{78}}, \bibinfo{pages}{034001}
  (\bibinfo{year}{2009}), \eprint{https://doi.org/10.1143/JPSJ.78.034001},
  \urlprefix\url{https://doi.org/10.1143/JPSJ.78.034001}.

\bibitem[{\citenamefont{Matson}(2012)}]{Matson:2012aa}
\bibinfo{author}{\bibfnamefont{J.}~\bibnamefont{Matson}},
  \bibinfo{journal}{Nature}  (\bibinfo{year}{2012}),
  \urlprefix\url{https://doi.org/10.1038/nature.2012.11163}.

\bibitem[{\citenamefont{Frey et~al.}(2014)\citenamefont{Frey, Funo, and
  Hotta}}]{Frey:2014}
\bibinfo{author}{\bibfnamefont{M.}~\bibnamefont{Frey}},
  \bibinfo{author}{\bibfnamefont{K.}~\bibnamefont{Funo}}, \bibnamefont{and}
  \bibinfo{author}{\bibfnamefont{M.}~\bibnamefont{Hotta}},
  \bibinfo{journal}{Phys. Rev. E} \textbf{\bibinfo{volume}{90}},
  \bibinfo{pages}{012127} (\bibinfo{year}{2014}),
  \urlprefix\url{https://link.aps.org/doi/10.1103/PhysRevE.90.012127}.

\bibitem[{\citenamefont{Ekert}(1991)}]{PhysRevLett.67.661}
\bibinfo{author}{\bibfnamefont{A.~K.} \bibnamefont{Ekert}},
  \bibinfo{journal}{Phys. Rev. Lett.} \textbf{\bibinfo{volume}{67}},
  \bibinfo{pages}{661} (\bibinfo{year}{1991}),
  \urlprefix\url{https://link.aps.org/doi/10.1103/PhysRevLett.67.661}.

\bibitem[{\citenamefont{Tittel et~al.}(1998)\citenamefont{Tittel, Brendel,
  Zbinden, and Gisin}}]{Tittel:1998ja}
\bibinfo{author}{\bibfnamefont{W.}~\bibnamefont{Tittel}},
  \bibinfo{author}{\bibfnamefont{J.}~\bibnamefont{Brendel}},
  \bibinfo{author}{\bibfnamefont{H.}~\bibnamefont{Zbinden}}, \bibnamefont{and}
  \bibinfo{author}{\bibfnamefont{N.}~\bibnamefont{Gisin}},
  \bibinfo{journal}{Phys. Rev. Lett.} \textbf{\bibinfo{volume}{81}},
  \bibinfo{pages}{3563} (\bibinfo{year}{1998}),
  \eprint{arXiv:quant-ph/9806043}.

\bibitem[{\citenamefont{Salart et~al.}(2008)\citenamefont{Salart, Baas,
  Branciard, Gisin, and Zbinden}}]{Salart-2008}
\bibinfo{author}{\bibfnamefont{D.}~\bibnamefont{Salart}},
  \bibinfo{author}{\bibfnamefont{A.}~\bibnamefont{Baas}},
  \bibinfo{author}{\bibfnamefont{C.}~\bibnamefont{Branciard}},
  \bibinfo{author}{\bibfnamefont{N.}~\bibnamefont{Gisin}}, \bibnamefont{and}
  \bibinfo{author}{\bibfnamefont{H.}~\bibnamefont{Zbinden}},
  \bibinfo{journal}{Nature} \textbf{\bibinfo{volume}{454}},
  \bibinfo{pages}{861} (\bibinfo{year}{2008}),
  \urlprefix\url{https://doi.org/10.1038/nature07121}.

\bibitem[{\citenamefont{Yin et~al.}(2020)\citenamefont{Yin, Li, Liao, Yang,
  Cao, Zhang, Ren, Cai, Liu, Li et~al.}}]{Yin:2020aa}
\bibinfo{author}{\bibfnamefont{J.}~\bibnamefont{Yin}},
  \bibinfo{author}{\bibfnamefont{Y.-H.} \bibnamefont{Li}},
  \bibinfo{author}{\bibfnamefont{S.-K.} \bibnamefont{Liao}},
  \bibinfo{author}{\bibfnamefont{M.}~\bibnamefont{Yang}},
  \bibinfo{author}{\bibfnamefont{Y.}~\bibnamefont{Cao}},
  \bibinfo{author}{\bibfnamefont{L.}~\bibnamefont{Zhang}},
  \bibinfo{author}{\bibfnamefont{J.-G.} \bibnamefont{Ren}},
  \bibinfo{author}{\bibfnamefont{W.-Q.} \bibnamefont{Cai}},
  \bibinfo{author}{\bibfnamefont{W.-Y.} \bibnamefont{Liu}},
  \bibinfo{author}{\bibfnamefont{S.-L.} \bibnamefont{Li}},
  \bibnamefont{et~al.}, \bibinfo{journal}{Nature}
  \textbf{\bibinfo{volume}{582}}, \bibinfo{pages}{501} (\bibinfo{year}{2020}),
  \urlprefix\url{https://doi.org/10.1038/s41586-020-2401-y}.

\bibitem[{\citenamefont{Mooney et~al.}(2019)\citenamefont{Mooney, Hill, and
  Hollenberg}}]{Mooney:2019aa}
\bibinfo{author}{\bibfnamefont{G.~J.} \bibnamefont{Mooney}},
  \bibinfo{author}{\bibfnamefont{C.~D.} \bibnamefont{Hill}}, \bibnamefont{and}
  \bibinfo{author}{\bibfnamefont{L.~C.~L.} \bibnamefont{Hollenberg}},
  \bibinfo{journal}{Scientific Reports} \textbf{\bibinfo{volume}{9}},
  \bibinfo{pages}{13465} (\bibinfo{year}{2019}),
  \urlprefix\url{https://doi.org/10.1038/s41598-019-49805-7}.

\bibitem[{\citenamefont{Summers and Werner}(1985)}]{SUMMERS1985257}
\bibinfo{author}{\bibfnamefont{S.~J.} \bibnamefont{Summers}} \bibnamefont{and}
  \bibinfo{author}{\bibfnamefont{R.}~\bibnamefont{Werner}},
  \bibinfo{journal}{Physics Letters A} \textbf{\bibinfo{volume}{110}},
  \bibinfo{pages}{257} (\bibinfo{year}{1985}), ISSN \bibinfo{issn}{0375-9601}.

\bibitem[{\citenamefont{Summers and
  Werner}(1987{\natexlab{a}})}]{Summers:1987ze}
\bibinfo{author}{\bibfnamefont{S.~J.} \bibnamefont{Summers}} \bibnamefont{and}
  \bibinfo{author}{\bibfnamefont{R.}~\bibnamefont{Werner}},
  \bibinfo{journal}{Commun. Math. Phys.} \textbf{\bibinfo{volume}{110}},
  \bibinfo{pages}{247} (\bibinfo{year}{1987}{\natexlab{a}}).

\bibitem[{\citenamefont{Summers and
  Werner}(1987{\natexlab{b}})}]{10.1063/1.527734}
\bibinfo{author}{\bibfnamefont{S.~J.} \bibnamefont{Summers}} \bibnamefont{and}
  \bibinfo{author}{\bibfnamefont{R.}~\bibnamefont{Werner}},
  \bibinfo{journal}{Journal of Mathematical Physics}
  \textbf{\bibinfo{volume}{28}}, \bibinfo{pages}{2448}
  (\bibinfo{year}{1987}{\natexlab{b}}), ISSN \bibinfo{issn}{0022-2488},
  \eprint{https://pubs.aip.org/aip/jmp/article-pdf/28/10/2448/11150126/2448\_1\_online.pdf},
  \urlprefix\url{https://doi.org/10.1063/1.527734}.

\bibitem[{\citenamefont{Summers and
  Werner}(1987{\natexlab{c}})}]{doi:10.1063/1.527733}
\bibinfo{author}{\bibfnamefont{S.~J.} \bibnamefont{Summers}} \bibnamefont{and}
  \bibinfo{author}{\bibfnamefont{R.}~\bibnamefont{Werner}},
  \bibinfo{journal}{Journal of Mathematical Physics}
  \textbf{\bibinfo{volume}{28}}, \bibinfo{pages}{2440}
  (\bibinfo{year}{1987}{\natexlab{c}}).

\bibitem[{\citenamefont{Peres and Terno}(2004)}]{RevModPhys.76.93}
\bibinfo{author}{\bibfnamefont{A.}~\bibnamefont{Peres}} \bibnamefont{and}
  \bibinfo{author}{\bibfnamefont{D.~R.} \bibnamefont{Terno}},
  \bibinfo{journal}{Rev. Mod. Phys.} \textbf{\bibinfo{volume}{76}},
  \bibinfo{pages}{93} (\bibinfo{year}{2004}),
  \urlprefix\url{https://link.aps.org/doi/10.1103/RevModPhys.76.93}.

\bibitem[{\citenamefont{HAWKING}(1974)}]{HAWKING:1974us}
\bibinfo{author}{\bibfnamefont{S.~W.} \bibnamefont{HAWKING}},
  \bibinfo{journal}{Nature} \textbf{\bibinfo{volume}{248}}, \bibinfo{pages}{30}
  (\bibinfo{year}{1974}), \urlprefix\url{https://doi.org/10.1038/248030a0}.

\bibitem[{\citenamefont{Hawking}(1975)}]{Hawking1975}
\bibinfo{author}{\bibfnamefont{S.~W.} \bibnamefont{Hawking}},
  \bibinfo{journal}{Communications in Mathematical Physics}
  \textbf{\bibinfo{volume}{43}}, \bibinfo{pages}{199} (\bibinfo{year}{1975}).

\bibitem[{\citenamefont{Unruh}(1976)}]{Unruh:1976db}
\bibinfo{author}{\bibfnamefont{W.~G.} \bibnamefont{Unruh}},
  \bibinfo{journal}{Phys. Rev. D} \textbf{\bibinfo{volume}{14}},
  \bibinfo{pages}{870} (\bibinfo{year}{1976}).

\bibitem[{\citenamefont{Birrell and Davies}(1982)}]{book:Birrell}
\bibinfo{author}{\bibfnamefont{N.~D.} \bibnamefont{Birrell}} \bibnamefont{and}
  \bibinfo{author}{\bibfnamefont{P.~C.~W.} \bibnamefont{Davies}},
  \emph{\bibinfo{title}{Quantum fields in curved space}}, Cambridge Monographs
  on Mathematical Physics (\bibinfo{publisher}{Cambridge University Press},
  \bibinfo{year}{1982}).

\bibitem[{\citenamefont{Unruh and Wald}(1984)}]{Unruh:1983ms}
\bibinfo{author}{\bibfnamefont{W.~G.} \bibnamefont{Unruh}} \bibnamefont{and}
  \bibinfo{author}{\bibfnamefont{R.~M.} \bibnamefont{Wald}},
  \bibinfo{journal}{Phys. Rev.} \textbf{\bibinfo{volume}{D29}},
  \bibinfo{pages}{1047} (\bibinfo{year}{1984}).

\bibitem[{\citenamefont{Crispino et~al.}(2008)\citenamefont{Crispino, Higuchi,
  and Matsas}}]{Crispino:2007eb}
\bibinfo{author}{\bibfnamefont{L.~C.~B.} \bibnamefont{Crispino}},
  \bibinfo{author}{\bibfnamefont{A.}~\bibnamefont{Higuchi}}, \bibnamefont{and}
  \bibinfo{author}{\bibfnamefont{G.~E.~A.} \bibnamefont{Matsas}},
  \bibinfo{journal}{Rev. Mod. Phys.} \textbf{\bibinfo{volume}{80}},
  \bibinfo{pages}{787} (\bibinfo{year}{2008}), \eprint{0710.5373}.

\bibitem[{\citenamefont{Takagi}(1986)}]{Takagi01031986}
\bibinfo{author}{\bibfnamefont{S.}~\bibnamefont{Takagi}},
  \bibinfo{journal}{Progress of Theoretical Physics Supplement}
  \textbf{\bibinfo{volume}{88}}, \bibinfo{pages}{1} (\bibinfo{year}{1986}),
  \eprint{http://ptps.oxfordjournals.org/content/88/1.full.pdf+html}.

\bibitem[{\citenamefont{Higuchi et~al.}(2017)\citenamefont{Higuchi, Iso, Ueda,
  and Yamamoto}}]{Higuchi:2017gcd}
\bibinfo{author}{\bibfnamefont{A.}~\bibnamefont{Higuchi}},
  \bibinfo{author}{\bibfnamefont{S.}~\bibnamefont{Iso}},
  \bibinfo{author}{\bibfnamefont{K.}~\bibnamefont{Ueda}}, \bibnamefont{and}
  \bibinfo{author}{\bibfnamefont{K.}~\bibnamefont{Yamamoto}},
  \bibinfo{journal}{Phys. Rev. D} \textbf{\bibinfo{volume}{96}},
  \bibinfo{pages}{083531} (\bibinfo{year}{2017}), \eprint{arXiv:1709.05757}.

\bibitem[{\citenamefont{Valentini}(1991)}]{VALENTINI1991321}
\bibinfo{author}{\bibfnamefont{A.}~\bibnamefont{Valentini}},
  \bibinfo{journal}{Physics Letters A} \textbf{\bibinfo{volume}{153}},
  \bibinfo{pages}{321} (\bibinfo{year}{1991}), ISSN \bibinfo{issn}{0375-9601}.

\bibitem[{\citenamefont{Reznik}(2003)}]{Reznik:2003uz}
\bibinfo{author}{\bibfnamefont{B.}~\bibnamefont{Reznik}},
  \bibinfo{journal}{Foundations of Physics} \textbf{\bibinfo{volume}{33}},
  \bibinfo{pages}{167} (\bibinfo{year}{2003}),
  \urlprefix\url{https://doi.org/10.1023/A:1022875910744}.

\bibitem[{\citenamefont{Reznik et~al.}(2005)\citenamefont{Reznik, Retzker, and
  Silman}}]{Reznik:2003mnx}
\bibinfo{author}{\bibfnamefont{B.}~\bibnamefont{Reznik}},
  \bibinfo{author}{\bibfnamefont{A.}~\bibnamefont{Retzker}}, \bibnamefont{and}
  \bibinfo{author}{\bibfnamefont{J.}~\bibnamefont{Silman}},
  \bibinfo{journal}{Phys. Rev. A} \textbf{\bibinfo{volume}{71}},
  \bibinfo{pages}{042104} (\bibinfo{year}{2005}),
  \eprint{arXiv:quant-ph/0310058}.

\bibitem[{\citenamefont{Koga et~al.}(2018)\citenamefont{Koga, Kimura, and
  Maeda}}]{Koga:2018the}
\bibinfo{author}{\bibfnamefont{J.-I.} \bibnamefont{Koga}},
  \bibinfo{author}{\bibfnamefont{G.}~\bibnamefont{Kimura}}, \bibnamefont{and}
  \bibinfo{author}{\bibfnamefont{K.}~\bibnamefont{Maeda}},
  \bibinfo{journal}{Phys. Rev. A} \textbf{\bibinfo{volume}{97}},
  \bibinfo{pages}{062338} (\bibinfo{year}{2018}), \eprint{arXiv:1804.01183}.

\bibitem[{\citenamefont{Ng et~al.}(2018)\citenamefont{Ng, Mann, and
  Mart\'\i{}n-Mart\'\i{}nez}}]{Ng:2018ilp}
\bibinfo{author}{\bibfnamefont{K.~K.} \bibnamefont{Ng}},
  \bibinfo{author}{\bibfnamefont{R.~B.} \bibnamefont{Mann}}, \bibnamefont{and}
  \bibinfo{author}{\bibfnamefont{E.}~\bibnamefont{Mart\'\i{}n-Mart\'\i{}nez}},
  \bibinfo{journal}{Phys. Rev. D} \textbf{\bibinfo{volume}{97}},
  \bibinfo{pages}{125011} (\bibinfo{year}{2018}), \eprint{arXiv:1805.01096}.

\bibitem[{\citenamefont{Koga et~al.}(2019)\citenamefont{Koga, Maeda, and
  Kimura}}]{Koga:2019fqh}
\bibinfo{author}{\bibfnamefont{J.-i.} \bibnamefont{Koga}},
  \bibinfo{author}{\bibfnamefont{K.}~\bibnamefont{Maeda}}, \bibnamefont{and}
  \bibinfo{author}{\bibfnamefont{G.}~\bibnamefont{Kimura}},
  \bibinfo{journal}{Phys. Rev. D} \textbf{\bibinfo{volume}{100}},
  \bibinfo{pages}{065013} (\bibinfo{year}{2019}), \eprint{1906.02843}.

\bibitem[{\citenamefont{Barman et~al.}(2021)\citenamefont{Barman, Barman, and
  Majhi}}]{Barman:2021bbw}
\bibinfo{author}{\bibfnamefont{D.}~\bibnamefont{Barman}},
  \bibinfo{author}{\bibfnamefont{S.}~\bibnamefont{Barman}}, \bibnamefont{and}
  \bibinfo{author}{\bibfnamefont{B.~R.} \bibnamefont{Majhi}},
  \bibinfo{journal}{JHEP} \textbf{\bibinfo{volume}{07}}, \bibinfo{pages}{124}
  (\bibinfo{year}{2021}), \eprint{arXiv:2104.11269}.

\bibitem[{\citenamefont{Barman et~al.}(2022{\natexlab{a}})\citenamefont{Barman,
  Barman, and Majhi}}]{Barman:2022xht}
\bibinfo{author}{\bibfnamefont{D.}~\bibnamefont{Barman}},
  \bibinfo{author}{\bibfnamefont{S.}~\bibnamefont{Barman}}, \bibnamefont{and}
  \bibinfo{author}{\bibfnamefont{B.~R.} \bibnamefont{Majhi}},
  \bibinfo{journal}{Phys. Rev. D} \textbf{\bibinfo{volume}{106}},
  \bibinfo{pages}{045005} (\bibinfo{year}{2022}{\natexlab{a}}),
  \eprint{2205.08505}.

\bibitem[{\citenamefont{Wu et~al.}(2024)\citenamefont{Wu, Wang, Huang, and
  Wang}}]{Wu:2024whx}
\bibinfo{author}{\bibfnamefont{S.-M.} \bibnamefont{Wu}},
  \bibinfo{author}{\bibfnamefont{R.-D.} \bibnamefont{Wang}},
  \bibinfo{author}{\bibfnamefont{X.-L.} \bibnamefont{Huang}}, \bibnamefont{and}
  \bibinfo{author}{\bibfnamefont{Z.}~\bibnamefont{Wang}}
  (\bibinfo{year}{2024}), \eprint{2408.11277}.

\bibitem[{\citenamefont{Pozas-Kerstjens and
  Martin-Martinez}(2016)}]{Pozas-Kerstjens:2016rsh}
\bibinfo{author}{\bibfnamefont{A.}~\bibnamefont{Pozas-Kerstjens}}
  \bibnamefont{and}
  \bibinfo{author}{\bibfnamefont{E.}~\bibnamefont{Martin-Martinez}},
  \bibinfo{journal}{Phys. Rev. D} \textbf{\bibinfo{volume}{94}},
  \bibinfo{pages}{064074} (\bibinfo{year}{2016}), \eprint{1605.07180}.

\bibitem[{\citenamefont{Perche et~al.}(2022)\citenamefont{Perche, Lima, and
  Mart\'\i{}n-Mart\'\i{}nez}}]{Perche:2021clp}
\bibinfo{author}{\bibfnamefont{T.~R.} \bibnamefont{Perche}},
  \bibinfo{author}{\bibfnamefont{C.}~\bibnamefont{Lima}}, \bibnamefont{and}
  \bibinfo{author}{\bibfnamefont{E.}~\bibnamefont{Mart\'\i{}n-Mart\'\i{}nez}},
  \bibinfo{journal}{Phys. Rev. D} \textbf{\bibinfo{volume}{105}},
  \bibinfo{pages}{065016} (\bibinfo{year}{2022}), \eprint{2111.12779}.

\bibitem[{\citenamefont{Liu et~al.}(2021)\citenamefont{Liu, Zhang, and
  Yu}}]{Liu:2020jaj}
\bibinfo{author}{\bibfnamefont{Z.}~\bibnamefont{Liu}},
  \bibinfo{author}{\bibfnamefont{J.}~\bibnamefont{Zhang}}, \bibnamefont{and}
  \bibinfo{author}{\bibfnamefont{H.}~\bibnamefont{Yu}}, \bibinfo{journal}{JHEP}
  \textbf{\bibinfo{volume}{08}}, \bibinfo{pages}{020} (\bibinfo{year}{2021}),
  \eprint{2101.00114}.

\bibitem[{\citenamefont{Barman and Majhi}(2023)}]{PhysRevD.108.085007}
\bibinfo{author}{\bibfnamefont{D.}~\bibnamefont{Barman}} \bibnamefont{and}
  \bibinfo{author}{\bibfnamefont{B.~R.} \bibnamefont{Majhi}},
  \bibinfo{journal}{Phys. Rev. D} \textbf{\bibinfo{volume}{108}},
  \bibinfo{pages}{085007} (\bibinfo{year}{2023}),
  \urlprefix\url{https://link.aps.org/doi/10.1103/PhysRevD.108.085007}.

\bibitem[{\citenamefont{Menezes}(2018)}]{Menezes:2017oeb}
\bibinfo{author}{\bibfnamefont{G.}~\bibnamefont{Menezes}},
  \bibinfo{journal}{Phys. Rev.} \textbf{\bibinfo{volume}{D97}},
  \bibinfo{pages}{085021} (\bibinfo{year}{2018}), \eprint{arXiv:1712.07151}.

\bibitem[{\citenamefont{Tjoa and Mann}(2020)}]{Tjoa:2020eqh}
\bibinfo{author}{\bibfnamefont{E.}~\bibnamefont{Tjoa}} \bibnamefont{and}
  \bibinfo{author}{\bibfnamefont{R.~B.} \bibnamefont{Mann}},
  \bibinfo{journal}{JHEP} \textbf{\bibinfo{volume}{08}}, \bibinfo{pages}{155}
  (\bibinfo{year}{2020}), \eprint{arXiv:2007.02955}.

\bibitem[{\citenamefont{Gallock-Yoshimura
  et~al.}(2021)\citenamefont{Gallock-Yoshimura, Tjoa, and
  Mann}}]{PhysRevD.104.025001}
\bibinfo{author}{\bibfnamefont{K.}~\bibnamefont{Gallock-Yoshimura}},
  \bibinfo{author}{\bibfnamefont{E.}~\bibnamefont{Tjoa}}, \bibnamefont{and}
  \bibinfo{author}{\bibfnamefont{R.~B.} \bibnamefont{Mann}},
  \bibinfo{journal}{Phys. Rev. D} \textbf{\bibinfo{volume}{104}},
  \bibinfo{pages}{025001} (\bibinfo{year}{2021}),
  \urlprefix\url{https://link.aps.org/doi/10.1103/PhysRevD.104.025001}.

\bibitem[{\citenamefont{Barman et~al.}(2022{\natexlab{b}})\citenamefont{Barman,
  Barman, and Majhi}}]{Barman:2021kwg}
\bibinfo{author}{\bibfnamefont{S.}~\bibnamefont{Barman}},
  \bibinfo{author}{\bibfnamefont{D.}~\bibnamefont{Barman}}, \bibnamefont{and}
  \bibinfo{author}{\bibfnamefont{B.~R.} \bibnamefont{Majhi}},
  \bibinfo{journal}{JHEP} \textbf{\bibinfo{volume}{09}}, \bibinfo{pages}{106}
  (\bibinfo{year}{2022}{\natexlab{b}}), \eprint{2112.01308}.

\bibitem[{\citenamefont{Barman and Majhi}(2024)}]{Barman:2023rhd}
\bibinfo{author}{\bibfnamefont{S.}~\bibnamefont{Barman}} \bibnamefont{and}
  \bibinfo{author}{\bibfnamefont{B.~R.} \bibnamefont{Majhi}},
  \bibinfo{journal}{Gen. Rel. Grav.} \textbf{\bibinfo{volume}{56}},
  \bibinfo{pages}{70} (\bibinfo{year}{2024}), \eprint{2301.06764}.

\bibitem[{\citenamefont{Kukita and Nambu}(2017)}]{Kukita:2017etu}
\bibinfo{author}{\bibfnamefont{S.}~\bibnamefont{Kukita}} \bibnamefont{and}
  \bibinfo{author}{\bibfnamefont{Y.}~\bibnamefont{Nambu}},
  \bibinfo{journal}{Entropy} \textbf{\bibinfo{volume}{19}},
  \bibinfo{pages}{449} (\bibinfo{year}{2017}), \eprint{arXiv:1708.01359}.

\bibitem[{\citenamefont{Martin-Martinez
  et~al.}(2016)\citenamefont{Martin-Martinez, Smith, and
  Terno}}]{Martin-Martinez:2015qwa}
\bibinfo{author}{\bibfnamefont{E.}~\bibnamefont{Martin-Martinez}},
  \bibinfo{author}{\bibfnamefont{A.~R.~H.} \bibnamefont{Smith}},
  \bibnamefont{and} \bibinfo{author}{\bibfnamefont{D.~R.} \bibnamefont{Terno}},
  \bibinfo{journal}{Phys. Rev. D} \textbf{\bibinfo{volume}{93}},
  \bibinfo{pages}{044001} (\bibinfo{year}{2016}), \eprint{arXiv:1507.02688}.

\bibitem[{\citenamefont{K et~al.}(2024)\citenamefont{K, Barman, and
  Kothawala}}]{K:2023oon}
\bibinfo{author}{\bibfnamefont{H.}~\bibnamefont{K}},
  \bibinfo{author}{\bibfnamefont{S.}~\bibnamefont{Barman}}, \bibnamefont{and}
  \bibinfo{author}{\bibfnamefont{D.}~\bibnamefont{Kothawala}},
  \bibinfo{journal}{Phys. Rev. D} \textbf{\bibinfo{volume}{109}},
  \bibinfo{pages}{065017} (\bibinfo{year}{2024}), \eprint{2311.15019}.

\bibitem[{\citenamefont{Ji et~al.}(2024)\citenamefont{Ji, Zhang, and
  Yu}}]{Ji:2024fcq}
\bibinfo{author}{\bibfnamefont{Y.}~\bibnamefont{Ji}},
  \bibinfo{author}{\bibfnamefont{J.}~\bibnamefont{Zhang}}, \bibnamefont{and}
  \bibinfo{author}{\bibfnamefont{H.}~\bibnamefont{Yu}}, \bibinfo{journal}{JHEP}
  \textbf{\bibinfo{volume}{06}}, \bibinfo{pages}{161} (\bibinfo{year}{2024}),
  \eprint{2401.13406}.

\bibitem[{\citenamefont{Lin and Mondal}(2024)}]{Lin:2024roh}
\bibinfo{author}{\bibfnamefont{F.-L.} \bibnamefont{Lin}} \bibnamefont{and}
  \bibinfo{author}{\bibfnamefont{S.}~\bibnamefont{Mondal}},
  \bibinfo{journal}{JHEP} \textbf{\bibinfo{volume}{08}}, \bibinfo{pages}{159}
  (\bibinfo{year}{2024}), \eprint{2406.19125}.

\bibitem[{\citenamefont{Mayank et~al.}(2025)\citenamefont{Mayank, Hari, Barman,
  and Kothawala}}]{Mayank:2025bkc}
\bibinfo{author}{\bibnamefont{Mayank}},
  \bibinfo{author}{\bibfnamefont{K.}~\bibnamefont{Hari}},
  \bibinfo{author}{\bibfnamefont{S.}~\bibnamefont{Barman}}, \bibnamefont{and}
  \bibinfo{author}{\bibfnamefont{D.}~\bibnamefont{Kothawala}}
  (\bibinfo{year}{2025}), \eprint{2502.20874}.

\bibitem[{\citenamefont{Barman et~al.}(2023)\citenamefont{Barman, Choudhury,
  Kad, and Majhi}}]{Barman:2022loh}
\bibinfo{author}{\bibfnamefont{D.}~\bibnamefont{Barman}},
  \bibinfo{author}{\bibfnamefont{A.}~\bibnamefont{Choudhury}},
  \bibinfo{author}{\bibfnamefont{B.}~\bibnamefont{Kad}}, \bibnamefont{and}
  \bibinfo{author}{\bibfnamefont{B.~R.} \bibnamefont{Majhi}},
  \bibinfo{journal}{Phys. Rev. D} \textbf{\bibinfo{volume}{107}},
  \bibinfo{pages}{045001} (\bibinfo{year}{2023}), \eprint{2211.00383}.

\bibitem[{\citenamefont{Chowdhury and Majhi}(2022)}]{Chowdhury:2021ieg}
\bibinfo{author}{\bibfnamefont{P.}~\bibnamefont{Chowdhury}} \bibnamefont{and}
  \bibinfo{author}{\bibfnamefont{B.~R.} \bibnamefont{Majhi}},
  \bibinfo{journal}{JHEP} \textbf{\bibinfo{volume}{05}}, \bibinfo{pages}{025}
  (\bibinfo{year}{2022}), \eprint{2110.11260}.

\bibitem[{\citenamefont{Nandi et~al.}(2024)\citenamefont{Nandi, Majhi, Debnath,
  and Kala}}]{Nandi:2024zxp}
\bibinfo{author}{\bibfnamefont{P.}~\bibnamefont{Nandi}},
  \bibinfo{author}{\bibfnamefont{B.~R.} \bibnamefont{Majhi}},
  \bibinfo{author}{\bibfnamefont{N.}~\bibnamefont{Debnath}}, \bibnamefont{and}
  \bibinfo{author}{\bibfnamefont{S.}~\bibnamefont{Kala}},
  \bibinfo{journal}{Phys. Lett. B} \textbf{\bibinfo{volume}{853}},
  \bibinfo{pages}{138706} (\bibinfo{year}{2024}), \eprint{2401.02778}.

\bibitem[{\citenamefont{Nandi and Majhi}(2024)}]{Nandi:2024jyf}
\bibinfo{author}{\bibfnamefont{P.}~\bibnamefont{Nandi}} \bibnamefont{and}
  \bibinfo{author}{\bibfnamefont{B.~R.} \bibnamefont{Majhi}},
  \bibinfo{journal}{Phys. Lett. B} \textbf{\bibinfo{volume}{857}},
  \bibinfo{pages}{138988} (\bibinfo{year}{2024}), \eprint{2403.11253}.

\bibitem[{\citenamefont{Dutta et~al.}(2025)\citenamefont{Dutta, Nandi, and
  Majhi}}]{Dutta:2025bge}
\bibinfo{author}{\bibfnamefont{M.}~\bibnamefont{Dutta}},
  \bibinfo{author}{\bibfnamefont{P.}~\bibnamefont{Nandi}}, \bibnamefont{and}
  \bibinfo{author}{\bibfnamefont{B.~R.} \bibnamefont{Majhi}}
  (\bibinfo{year}{2025}), \eprint{2503.19688}.

\bibitem[{\citenamefont{Pozas-Kerstjens
  et~al.}(2017)\citenamefont{Pozas-Kerstjens, Louko, and
  Martin-Martinez}}]{Pozas-Kerstjens:2017xjr}
\bibinfo{author}{\bibfnamefont{A.}~\bibnamefont{Pozas-Kerstjens}},
  \bibinfo{author}{\bibfnamefont{J.}~\bibnamefont{Louko}}, \bibnamefont{and}
  \bibinfo{author}{\bibfnamefont{E.}~\bibnamefont{Martin-Martinez}},
  \bibinfo{journal}{Phys. Rev. D} \textbf{\bibinfo{volume}{95}},
  \bibinfo{pages}{105009} (\bibinfo{year}{2017}), \eprint{1703.02982}.

\bibitem[{\citenamefont{Henderson and Menicucci}(2020)}]{Henderson:2020ucx}
\bibinfo{author}{\bibfnamefont{L.~J.} \bibnamefont{Henderson}}
  \bibnamefont{and} \bibinfo{author}{\bibfnamefont{N.~C.}
  \bibnamefont{Menicucci}}, \bibinfo{journal}{Phys. Rev. D}
  \textbf{\bibinfo{volume}{102}}, \bibinfo{pages}{125026}
  (\bibinfo{year}{2020}), \eprint{arXiv:2005.05330}.

\bibitem[{\citenamefont{Shallue and Carroll}(2025)}]{Shallue:2025zto}
\bibinfo{author}{\bibfnamefont{C.~J.} \bibnamefont{Shallue}} \bibnamefont{and}
  \bibinfo{author}{\bibfnamefont{S.~M.} \bibnamefont{Carroll}}
  (\bibinfo{year}{2025}), \eprint{2501.06609}.

\bibitem[{\citenamefont{Stargen}(2025)}]{Stargen:2025prb}
\bibinfo{author}{\bibfnamefont{D.~J.} \bibnamefont{Stargen}}
  (\bibinfo{year}{2025}), \eprint{2501.09676}.

\bibitem[{\citenamefont{Lochan and Padmanabhan}(2025)}]{Lochan:2025mru}
\bibinfo{author}{\bibfnamefont{K.}~\bibnamefont{Lochan}} \bibnamefont{and}
  \bibinfo{author}{\bibfnamefont{T.}~\bibnamefont{Padmanabhan}},
  \bibinfo{journal}{Class. Quant. Grav.} \textbf{\bibinfo{volume}{42}},
  \bibinfo{pages}{03LT01} (\bibinfo{year}{2025}).

\bibitem[{\citenamefont{Sahota and Lochan}(2025)}]{PhysRevD.111.045023}
\bibinfo{author}{\bibfnamefont{H.~S.} \bibnamefont{Sahota}} \bibnamefont{and}
  \bibinfo{author}{\bibfnamefont{K.}~\bibnamefont{Lochan}},
  \bibinfo{journal}{Phys. Rev. D} \textbf{\bibinfo{volume}{111}},
  \bibinfo{pages}{045023} (\bibinfo{year}{2025}),
  \urlprefix\url{https://link.aps.org/doi/10.1103/PhysRevD.111.045023}.

\bibitem[{\citenamefont{Gutti et~al.}(2023)\citenamefont{Gutti, Nair, and
  Samantray}}]{Gutti:2022xov}
\bibinfo{author}{\bibfnamefont{S.}~\bibnamefont{Gutti}},
  \bibinfo{author}{\bibfnamefont{A.~U.} \bibnamefont{Nair}}, \bibnamefont{and}
  \bibinfo{author}{\bibfnamefont{P.}~\bibnamefont{Samantray}},
  \bibinfo{journal}{Phys. Rev. D} \textbf{\bibinfo{volume}{108}},
  \bibinfo{pages}{025010} (\bibinfo{year}{2023}), \eprint{2210.08925}.

\bibitem[{\citenamefont{Nair et~al.}(2024)\citenamefont{Nair, Jha, Samantray,
  and Gutti}}]{Nair:2024ryr}
\bibinfo{author}{\bibfnamefont{A.~U.} \bibnamefont{Nair}},
  \bibinfo{author}{\bibfnamefont{R.~K.} \bibnamefont{Jha}},
  \bibinfo{author}{\bibfnamefont{P.}~\bibnamefont{Samantray}},
  \bibnamefont{and} \bibinfo{author}{\bibfnamefont{S.}~\bibnamefont{Gutti}}
  (\bibinfo{year}{2024}), \eprint{2412.02560}.

\bibitem[{\citenamefont{Michael A.~{Nielsen}}(2004)}]{book:nielsen}
\bibinfo{author}{\bibfnamefont{I.~L.~C.} \bibnamefont{Michael A.~{Nielsen}}},
  \emph{\bibinfo{title}{Quantum computation and quantum information}},
  Cambridge Series on Information and the Natural Sciences
  (\bibinfo{publisher}{Cambridge University Press}, \bibinfo{year}{2004}),
  \bibinfo{edition}{1st} ed., ISBN
  \bibinfo{isbn}{0521635039,9780521635035,0521632358},
  \urlprefix\url{http://gen.lib.rus.ec/book/index.php?md5=2badbdeda09e90d88305d6da93094849}.

\bibitem[{\citenamefont{Simidzija and
  Mart\'\i{}n-Mart\'\i{}nez}(2018)}]{Simidzija:2018ddw}
\bibinfo{author}{\bibfnamefont{P.}~\bibnamefont{Simidzija}} \bibnamefont{and}
  \bibinfo{author}{\bibfnamefont{E.}~\bibnamefont{Mart\'\i{}n-Mart\'\i{}nez}},
  \bibinfo{journal}{Phys. Rev. D} \textbf{\bibinfo{volume}{98}},
  \bibinfo{pages}{085007} (\bibinfo{year}{2018}), \eprint{arXiv:1809.05547}.

\bibitem[{\citenamefont{Ollivier and Zurek}(2001)}]{PhysRevLett.88.017901}
\bibinfo{author}{\bibfnamefont{H.}~\bibnamefont{Ollivier}} \bibnamefont{and}
  \bibinfo{author}{\bibfnamefont{W.~H.} \bibnamefont{Zurek}},
  \bibinfo{journal}{Phys. Rev. Lett.} \textbf{\bibinfo{volume}{88}},
  \bibinfo{pages}{017901} (\bibinfo{year}{2001}),
  \urlprefix\url{https://link.aps.org/doi/10.1103/PhysRevLett.88.017901}.

\bibitem[{\citenamefont{Henderson and Vedral}(2001)}]{Henderson_2001}
\bibinfo{author}{\bibfnamefont{L.}~\bibnamefont{Henderson}} \bibnamefont{and}
  \bibinfo{author}{\bibfnamefont{V.}~\bibnamefont{Vedral}},
  \bibinfo{journal}{Journal of Physics A: Mathematical and General}
  \textbf{\bibinfo{volume}{34}}, \bibinfo{pages}{6899} (\bibinfo{year}{2001}),
  ISSN \bibinfo{issn}{1361-6447},
  \urlprefix\url{http://dx.doi.org/10.1088/0305-4470/34/35/315}.

\bibitem[{\citenamefont{Sarandy}(2009)}]{PhysRevA.80.022108}
\bibinfo{author}{\bibfnamefont{M.~S.} \bibnamefont{Sarandy}},
  \bibinfo{journal}{Phys. Rev. A} \textbf{\bibinfo{volume}{80}},
  \bibinfo{pages}{022108} (\bibinfo{year}{2009}),
  \urlprefix\url{https://link.aps.org/doi/10.1103/PhysRevA.80.022108}.

\bibitem[{\citenamefont{Luo}(2008)}]{PhysRevA.77.042303}
\bibinfo{author}{\bibfnamefont{S.}~\bibnamefont{Luo}}, \bibinfo{journal}{Phys.
  Rev. A} \textbf{\bibinfo{volume}{77}}, \bibinfo{pages}{042303}
  (\bibinfo{year}{2008}),
  \urlprefix\url{https://link.aps.org/doi/10.1103/PhysRevA.77.042303}.

\bibitem[{\citenamefont{Ali et~al.}(2010)\citenamefont{Ali, Rau, and
  Alber}}]{PhysRevA.81.042105}
\bibinfo{author}{\bibfnamefont{M.}~\bibnamefont{Ali}},
  \bibinfo{author}{\bibfnamefont{A.~R.~P.} \bibnamefont{Rau}},
  \bibnamefont{and} \bibinfo{author}{\bibfnamefont{G.}~\bibnamefont{Alber}},
  \bibinfo{journal}{Phys. Rev. A} \textbf{\bibinfo{volume}{81}},
  \bibinfo{pages}{042105} (\bibinfo{year}{2010}),
  \urlprefix\url{https://link.aps.org/doi/10.1103/PhysRevA.81.042105}.

\bibitem[{\citenamefont{Chen et~al.}(2011)\citenamefont{Chen, Zhang, Yu, Yi,
  and Oh}}]{PhysRevA.84.042313}
\bibinfo{author}{\bibfnamefont{Q.}~\bibnamefont{Chen}},
  \bibinfo{author}{\bibfnamefont{C.}~\bibnamefont{Zhang}},
  \bibinfo{author}{\bibfnamefont{S.}~\bibnamefont{Yu}},
  \bibinfo{author}{\bibfnamefont{X.~X.} \bibnamefont{Yi}}, \bibnamefont{and}
  \bibinfo{author}{\bibfnamefont{C.~H.} \bibnamefont{Oh}},
  \bibinfo{journal}{Phys. Rev. A} \textbf{\bibinfo{volume}{84}},
  \bibinfo{pages}{042313} (\bibinfo{year}{2011}),
  \urlprefix\url{https://link.aps.org/doi/10.1103/PhysRevA.84.042313}.

\bibitem[{\citenamefont{Pal and Bose}(2011)}]{Pal_2011}
\bibinfo{author}{\bibfnamefont{A.~K.} \bibnamefont{Pal}} \bibnamefont{and}
  \bibinfo{author}{\bibfnamefont{I.}~\bibnamefont{Bose}},
  \bibinfo{journal}{Journal of Physics B: Atomic, Molecular and Optical
  Physics} \textbf{\bibinfo{volume}{44}}, \bibinfo{pages}{045101}
  (\bibinfo{year}{2011}), ISSN \bibinfo{issn}{1361-6455},
  \urlprefix\url{http://dx.doi.org/10.1088/0953-4075/44/4/045101}.

\bibitem[{\citenamefont{Huang}(2013)}]{PhysRevA.88.014302}
\bibinfo{author}{\bibfnamefont{Y.}~\bibnamefont{Huang}},
  \bibinfo{journal}{Phys. Rev. A} \textbf{\bibinfo{volume}{88}},
  \bibinfo{pages}{014302} (\bibinfo{year}{2013}),
  \urlprefix\url{https://link.aps.org/doi/10.1103/PhysRevA.88.014302}.

\bibitem[{\citenamefont{Yurischev}(2015)}]{Yurischev_2015}
\bibinfo{author}{\bibfnamefont{M.~A.} \bibnamefont{Yurischev}},
  \bibinfo{journal}{Quantum Information Processing}
  \textbf{\bibinfo{volume}{14}}, \bibinfo{pages}{3399} (\bibinfo{year}{2015}),
  ISSN \bibinfo{issn}{1573-1332},
  \urlprefix\url{http://dx.doi.org/10.1007/s11128-015-1046-5}.

\bibitem[{\citenamefont{Guo}(2016)}]{YuGuo_2016}
\bibinfo{author}{\bibfnamefont{Y.}~\bibnamefont{Guo}},
  \bibinfo{journal}{Scientific Reports} \textbf{\bibinfo{volume}{6}},
  \bibinfo{pages}{25241} (\bibinfo{year}{2016}),
  \urlprefix\url{https://doi.org/10.1038/srep25241}.

\bibitem[{\citenamefont{Riedel~G{\aa}rding
  et~al.}(2021)\citenamefont{Riedel~G{\aa}rding, Schwaller, Chan, Chang, Bosch,
  Gessler, Laborde, Hernandez, Si, Dupertuis et~al.}}]{e23070797}
\bibinfo{author}{\bibfnamefont{E.}~\bibnamefont{Riedel~G{\aa}rding}},
  \bibinfo{author}{\bibfnamefont{N.}~\bibnamefont{Schwaller}},
  \bibinfo{author}{\bibfnamefont{C.~L.} \bibnamefont{Chan}},
  \bibinfo{author}{\bibfnamefont{S.~Y.} \bibnamefont{Chang}},
  \bibinfo{author}{\bibfnamefont{S.}~\bibnamefont{Bosch}},
  \bibinfo{author}{\bibfnamefont{F.}~\bibnamefont{Gessler}},
  \bibinfo{author}{\bibfnamefont{W.~R.} \bibnamefont{Laborde}},
  \bibinfo{author}{\bibfnamefont{J.~N.} \bibnamefont{Hernandez}},
  \bibinfo{author}{\bibfnamefont{X.}~\bibnamefont{Si}},
  \bibinfo{author}{\bibfnamefont{M.-A.} \bibnamefont{Dupertuis}},
  \bibnamefont{et~al.}, \bibinfo{journal}{Entropy}
  \textbf{\bibinfo{volume}{23}} (\bibinfo{year}{2021}), ISSN
  \bibinfo{issn}{1099-4300},
  \urlprefix\url{https://www.mdpi.com/1099-4300/23/7/797}.

\bibitem[{\citenamefont{Mart\'{\i}n-Mart\'{\i}nez and
  Rodriguez-Lopez}(2018)}]{PhysRevD.97.105026}
\bibinfo{author}{\bibfnamefont{E.}~\bibnamefont{Mart\'{\i}n-Mart\'{\i}nez}}
  \bibnamefont{and}
  \bibinfo{author}{\bibfnamefont{P.}~\bibnamefont{Rodriguez-Lopez}},
  \bibinfo{journal}{Phys. Rev. D} \textbf{\bibinfo{volume}{97}},
  \bibinfo{pages}{105026} (\bibinfo{year}{2018}),
  \urlprefix\url{https://link.aps.org/doi/10.1103/PhysRevD.97.105026}.

\bibitem[{\citenamefont{Peres}(1996)}]{Peres:1996dw}
\bibinfo{author}{\bibfnamefont{A.}~\bibnamefont{Peres}},
  \bibinfo{journal}{Phys. Rev. Lett.} \textbf{\bibinfo{volume}{77}},
  \bibinfo{pages}{1413} (\bibinfo{year}{1996}),
  \eprint{arXiv:quant-ph/9604005}.

\bibitem[{\citenamefont{Horodecki et~al.}(1996)\citenamefont{Horodecki,
  Horodecki, and Horodecki}}]{Horodecki:1996nc}
\bibinfo{author}{\bibfnamefont{M.}~\bibnamefont{Horodecki}},
  \bibinfo{author}{\bibfnamefont{P.}~\bibnamefont{Horodecki}},
  \bibnamefont{and}
  \bibinfo{author}{\bibfnamefont{R.}~\bibnamefont{Horodecki}},
  \bibinfo{journal}{Phys. Lett. A} \textbf{\bibinfo{volume}{223}},
  \bibinfo{pages}{1} (\bibinfo{year}{1996}), \eprint{arXiv:quant-ph/9605038}.

\bibitem[{\citenamefont{Bennett et~al.}(1996)\citenamefont{Bennett, DiVincenzo,
  Smolin, and Wootters}}]{Bennett:1996gf}
\bibinfo{author}{\bibfnamefont{C.~H.} \bibnamefont{Bennett}},
  \bibinfo{author}{\bibfnamefont{D.~P.} \bibnamefont{DiVincenzo}},
  \bibinfo{author}{\bibfnamefont{J.~A.} \bibnamefont{Smolin}},
  \bibnamefont{and} \bibinfo{author}{\bibfnamefont{W.~K.}
  \bibnamefont{Wootters}}, \bibinfo{journal}{Phys. Rev. A}
  \textbf{\bibinfo{volume}{54}}, \bibinfo{pages}{3824} (\bibinfo{year}{1996}),
  \eprint{arXiv:quant-ph/9604024}.

\bibitem[{\citenamefont{Hill and Wootters}(1997)}]{Hill:1997pfa}
\bibinfo{author}{\bibfnamefont{S.}~\bibnamefont{Hill}} \bibnamefont{and}
  \bibinfo{author}{\bibfnamefont{W.~K.} \bibnamefont{Wootters}},
  \bibinfo{journal}{Phys. Rev. Lett.} \textbf{\bibinfo{volume}{78}},
  \bibinfo{pages}{5022} (\bibinfo{year}{1997}),
  \eprint{arXiv:quant-ph/9703041}.

\bibitem[{\citenamefont{Wootters}(1998)}]{Wootters:1997id}
\bibinfo{author}{\bibfnamefont{W.~K.} \bibnamefont{Wootters}},
  \bibinfo{journal}{Phys. Rev. Lett.} \textbf{\bibinfo{volume}{80}},
  \bibinfo{pages}{2245} (\bibinfo{year}{1998}),
  \eprint{arXiv:quant-ph/9709029}.

\bibitem[{\citenamefont{Oberhettinger}(1990)}]{TablesFT}
\bibinfo{author}{\bibfnamefont{F.}~\bibnamefont{Oberhettinger}},
  \emph{\bibinfo{title}{Tables of Fourier Transforms and Fourier Transforms of
  Distributions}}, vol. \bibinfo{volume}{pp. 204, Chap. 3,
  10.1007/978-3-642-74349-8} (\bibinfo{publisher}{Springer, Berlin,
  Heidelberg}, \bibinfo{year}{1990}).

\bibitem[{{\relax DLMF}()}]{NIST:DLMF}
{\relax DLMF}, \emph{\bibinfo{title}{{\it NIST Digital Library of Mathematical
  Functions}}}, \bibinfo{note}{f.~W.~J. Olver, A.~B. {Olde Daalhuis}, D.~W.
  Lozier, B.~I. Schneider, R.~F. Boisvert, C.~W. Clark, B.~R. Miller and B.~V.
  Saunders, eds.}

\end{thebibliography}

\end{document}